\documentclass[
  journal=pasa,
  manuscript=research-paper, 
  year=2020,
  volume=37,
]{cup-journal}

\usepackage{microtype,siunitx,booktabs}
\usepackage{listings} 
\usepackage[most]{tcolorbox}
\usepackage{inconsolata}
\newtcblisting[auto counter]{sexylisting}[2][]{sharp corners, 
    fonttitle=\bfseries, colframe=gray, listing only, 
    listing options={basicstyle=\ttfamily,language=C++,breaklines=true}, 
    title=Listing \thetcbcounter: #2, #1}

\usepackage{comment}
\usepackage{amsmath,amssymb} 
\usepackage{graphicx} 
\usepackage{caption,subcaption}
\usepackage{geometry}

\usepackage{gensymb} 
\usepackage{xfrac}

\usepackage{caption}
\usepackage{subcaption}

\usepackage{tabularx}
\usepackage{booktabs}

\usepackage[colorlinks=true,allcolors=blue]{hyperref}%

\title{High-Time Resolution GPU Imager for FRB searches at low radio frequencies}

\author{M.~Sokolowski}
\affiliation{International Centre for Radio Astronomy Research, Curtin University, Bentley, WA6102, Australia}
\email[M.~Sokolowski]{marcin.sokolowski@curtin.edu.au}

\author{G.~Aniruddha}
\affiliation{International Centre for Radio Astronomy Research, Curtin University, Bentley, WA6102, Australia}

\author{C.~Di~Pietrantonio}
\affiliation{Curtin University, Pawsey Supercomputing Research Centre, Kensington WA6151}

\author{C.~Harris}
\affiliation{Pawsey Supercomputing Research Centre, Kensington WA6151}

\author{D.~C.~Price}
\affiliation{International Centre for Radio Astronomy Research, Curtin University, Bentley, WA6102, Australia}
\affiliation{Square Kilometre Array Observatory (SKAO), Kensington WA6151}

\author{S.~McSweeney}
\affiliation{International Centre for Radio Astronomy Research, Curtin University, Bentley, WA6102, Australia}

\author{R.~B.~Wayth}
\affiliation{International Centre for Radio Astronomy Research, Curtin University, Bentley, WA6102, Australia}

\author{N.~D.~R. Bhat}
\affiliation{International Centre for Radio Astronomy Research, Curtin University, Bentley, WA6102, Australia}

\doi{10.1017/pasa.2020.32}

\received {31 May 2023}
\revised  {dd Mmm 2023}
\accepted {dd Mmm 2023}
\published{dd Mmm 2023}


\begin{document}

\emergencystretch 3em



\newcommand{\todo}[1] {#1}
\newcommand{\red}[1] {#1}

\begin{abstract}
Fast Radio Bursts (FRBs) are millisecond dispersed radio pulses of predominately extra-galactic origin. Although originally discovered at GHz frequencies, most FRBs have been detected between 400 to 800\,MHz. Nevertheless, only a handful of FRBs were detected at radio frequencies $\le$400\,MHz. Searching for FRBs at low frequencies is computationally challenging due to increased dispersive delay that must be accounted for. Nevertheless, the wide field of view (FoV) of low-frequency telescopes -- such as the the Murchison Widefield Array (MWA), and prototype stations of the low-frequency Square Kilometre Array (SKA-Low) -- makes them promising instruments to open a low-frequency window on FRB event rates, and constrain FRB emission models. The standard approach, inherited from high-frequencies, is to form multiple tied-array beams to tessellate the entire FoV and perform the search on the resulting time series. This approach, however, may not be optimal for low-frequency interferometers due to their large FoVs and high spatial resolutions leading to a large number of beams. Consequently, there are regions of parameter space in terms of number of antennas and resolution elements (pixels) where interferometric imaging is computationally more efficient. Here we present a new high-time resolution imager \emph{BLINK} implemented on modern Graphical Processing Units (GPUs) and intended for radio astronomy data. The main goal for this imager is to become part of a fully GPU-accelerated FRB search pipeline. We describe the imager and present its verification on real and simulated data processed to form all-sky and widefield images from the MWA and prototype SKA-Low stations. We also present and compare benchmarks of the GPU and CPU code executed on laptops, desktop computers, and Australian supercomputers. The code is publicly available at \url{https://github.com/PaCER-BLINK-Project/imager} and
can be applied to data from any radio telescope.
\end{abstract}

\section{Introduction}
\label{sec_introduction}

\red{Fast Radio Bursts (FRBs) are extremely interesting millisecond duration radio pulses \citep[recent reviews in ][]{2022A&ARv..30....2P,2021Univ....8....9P,2019ARA&A..57..417C} with flux densities and cosmological redshifts implying huge energies ($\sim$10$^{39}$ erg). The first \red{FRB 20010724A}, also known as Lorimer Burst, was discovered in 2007 \citep{2007Sci...318..777L}. The subsequent detections by \citet{2013Sci...341...53T}, and the discovery of the first repeating \red{FRB 20121102A} \citep{2014ApJ...790..101S} at redshift z$\approx$0.19  \citep{2017ApJ...834L...7T} established FRBs as a new astrophysical phenomena. In the following years, the Commensal Realtime ASKAP Fast Transients (CRAFT) survey at 1.4\,GHz \citep{Macquartetal2010} discovered and localised multiple FRBs \citep[e.g.][]{2019Sci...365..565B,2019Sci...366..231P,2020ApJ...895L..37B} including the most distant FRB at the redshift of $\sim 1$ \citep{2023Sci...382..294R}. FRBs were also demonstrated to be very precise direct probes of baryonic matter on cosmological scales \citep[e.g. ][]{2020Natur.581..391M,2022MNRAS.516.4862J}. }

\red{Despite the growing observational evidence the physical mechanisms powering FRBs remain unexplained (see \citet{2019A&ARv..27....4P,2022A&ARv..30....2P} or FRB Theory Catalogue\footnote{\url{https://frbtheorycat.org/index.php}}). Improving understanding of progenitors and physical processes behind FRBs requires broadband and multi-wavelength detections \citep{2021Univ....7...76N}. However, except a single case of the Galactic magnetar Galactic Soft Gamma Repeater SGR 1935+2154 \citep{2020Natur.587...59B,2020Natur.587...54C}, no FRB was detected at electromagnetic wavelengths other than radio. The desired broadband radio detections and multi-wavelength observations can be achieved by targeting bright FRBs from the local Universe \citep[see ][to name a few]{2019MNRAS.490....1A,2024MNRAS.527.3659D,2022Natur.602..585K}. As discussed by \citet{2021Univ....8....9P}, detections of nearby FRBs  (z$\lesssim$0.5) at frequencies below 400\,MHz can be achieved by small arrays with large field of view (FoV) and dedicated on-sky time. For example, an all-sky transient monitoring systems \citep[e.g.][]{2022aapr.confE...1S,2021PASA...38...23S} can robustly measure low-frequency FRB rate and improve our understanding of the FRB population. Furthermore, simultaneous detections at low and high frequencies can provide important input for broad-band spectral modelling to study FRB progenitors, emission mechanisms and constrain their energies (depending on spectral characteristics, e.g. the presence a low frequency cut-off). }

Since, the very beginning low-frequency interferometers such as the Murchison Widefield Array \citep[MWA;][]{2013PASA...30....7T,2018PASA...35...33W} and the LOw-Frequency ARray \citep[LOFAR;][]{2013A&A...556A...2V}
tried to detect FRBs at frequencies below 300\,MHz. Non-targeted (also known as ``blind'') searches for low-frequency FRBs  were conducted either using tied-array beamforming, standard pulsar search packages like PRESTO~\citep[][]{2011ascl.soft07017R} or image-based approaches. \citet{2014A&A...570A..60C} performed an FRB search in nearly 300\,hours of incoherently and coherently beamformed LOFAR data and did not detect any FRBs down to fluence threshold of $\approx$71\,Jy\,ms, while \citet{2015MNRAS.452.1254K} did not detect FRBs down to about 310\,Jy\,ms in 1446 hours of data. Beamformed searches were also performed with the Long Wavelength Array (LWA), and resulted in detection of radio transients of unknown origin \citep{2019ApJ...874..151V} but, so far, no FRBs \citep{2019ApJ...886..123A}. Most of the efforts with the MWA were undertaken in the image domain. In the early searches, \citet{2015AJ....150..199T} analysed 10.5\,hours of MWA observations in 2\,s time resolution but did not detect any FRBs above fluence threshold of 700\,Jy\,ms. In a similar search using 100\,hours of 28\,s images, \citet{2016MNRAS.458.3506R} did not detect any FRB down to fluence 7980\,Jy\,ms. \citet{2018ApJ...867L..12S} co-observed the same fields as ASKAP CRAFT, and no low-frequency counterparts of the seven ASKAP FRBs were found in the simultaneously recorded MWA data with the stringiest limit of 450\,Jy\,ms for \red{FRB 20180324A}. More recently, \citet{2023MNRAS.518.4278T} performed targeted  search for low-frequency FRB-like signals from known repeating FRBs and short Gamma-Ray Bursts (GRBs) in beamformed MWA data, but none were found \citep[][]{2022MNRAS.514.2756T,2022PASA...39....3T}. It is worth highlighting that non-targeted searches were unsuccessful mainly due to relatively small amount of processed data, which was limited by the efficiency of the available software packages. \red{Therefore, it is imperative to develop efficient software packages to boost the amount of processed data to many thousands of hours.}

Since 2018, the Canadian Hydrogen Intensity Mapping Experiment \citep[CHIME/FRB;][]{Amiri_2018} detected hundreds of FRBs in the frequency band 400 -- 800\,MHz, and signals from many of these FRBs were observed down to 400\,MHz indicating that at least some FRBs can be observed at even lower frequencies. This was ultimately confirmed when LOFAR detected pulses from the CHIME repeating \red{FRB 20180916B} \citep{2021ApJ...911L...3P,2021Natur.596..505P}, and Green Bank Telescope (GBT) detected FRB 20200125A at 350\,MHz \citep{2020ApJ...904...92P}. \red{The observed pulse widths of these FRBs ($\sim$10 -- 100\,ms) indicate that moderate time resolutions ($\sim$50\,ms) may be sufficient to detect FRBs at these frequencies.}

Hence, there is a growing evidence that at least some FRBs can be observed at frequencies $\le$350\,MHz. The main reasons for a very small number of low-frequency detections are likely: smaller number of events due to physical mechanisms (absorption and pulse broadening due to scattering), limited on-sky time and higher computational complexity of the searches (see discussion in Section~\ref{sect_frb_search_methods}). The presented high-time resolution GPU imager is a step towards addressing at least the latter two of these issues, and making image-based FRB searches with low-frequency interferometers computationally feasible, affordable and, hopefully, successful. This is supported by the fact, that in some parts of parameter space (number of antennas above 100) computational cost of interferometric imaging can be even order of magnitude lower than cost of the commonly used tied-array beamforming (see Section~\ref{subsec_comp_cost}).

One of the reasons for implementing a new software package is that many existing software pipelines are tightly coupled with the instruments they were programmed for and cannot be easily adopted. Hence, one of the aims of this project is to make the software publicly available and applicable to data from many radio telescopes. In order to achieve this, the I/O layer has been separated into a dedicated library where telescope-specific I/O functions can be implemented. 

The GPU hardware market becomes more diversified which drives the hardware prices lower. This leads to increasing contribution of GPUs to the computational power of High-Performance Computing (HPC) centres. In particular, the new supercomputer Setonix at the Pawsey Supercomputing Centre (Pawsey)\footnote{\url{https://pawsey.org.au/}} has the total (including CPUs and GPUs) peak performance power of the order of 43\,petaFlops\footnote{\url{https://discover.pawsey.org.au/project/setonix}} with approximately 80\% of this (about 35\,petaFlops\footnote{\url{https://www.top500.org/lists/top500/list/2022/11/}}) provided by GPUs. Moreover, the peak performance of 57 GFlops/Watt makes Setonix the 4$^{th}$ greenest supercomputer in the world\footnote{\url{https://pawsey.org.au/australias-setonix-ranking/}}. The carbon footprint of astronomy infrastructure and computing centres is becoming increasingly important issue for the astronomy community \citep{2022NatAs...6..503K}, which is another factor strongly supporting the transition to GPU-based computing. In-line with this, energy efficiency is becoming a new metrics in the new accounting models for heterogeneous supercomputers \citep{2021arXiv211009987D}, and GPUs can be even an order of magnitude more energy efficient for complex computational problems \citep{8782524}. In radio astronomy context, NVIDIA Tensor Cores were shown to be about 5 to 10 times faster in correlator applications and in the same time 5 to 10 times more energy efficient than normal GPU cores \citep{2021A&A...656A..52R}. 

Nevertheless, despite all the above, most of the existing radio astronomy software was developed for either CPUs or, \red{in very few cases}, specifically for NVIDIA GPUs (cannot be used with AMD hardware). Therefore, it is increasingly important to develop energy efficient GPU-based software for radio astronomy, which is one of the over-arching goals of this project. Given the importance of GPUs for the efficiency of Setonix, Pawsey initiated PaCER projects\footnote{\url{https://pawsey.org.au/pacer/}} to convert existing software for various research applications or develop new software packages suitable for both AMD and NVIDIA GPUs. The development of the presented high-time resolution GPU imager was also supported by the PaCER initiative. 

The remainder of this paper is organised as follows. In Section~\ref{sect_frb_search_methods} we provide a brief summary of the FRB search methods used by low-frequency interferometers. In Section~\ref{sec_target_instruments} we describe the primary target instruments for the developed GPU imaging software. Section~\ref{sec_radioastronomy_imaging} provides a short overview of the standard interferometric imaging process, while Section~\ref{sec_implementation} describes the design, implementation, and validation of the CPU and GPU version of the presented high-time resolution imager. Section~\ref{sec_benchmarking} summarises the results of imager's benchmarking on various compute and GPU architectures. Finally, Section~\ref{sec_summary} summarises the work and discusses future plans.

\section{FRB Search Methods}
\label{sect_frb_search_methods}

In the light of hundreds of FRBs detected by CHIME down to 400\,MHz, the small number of detections by interferometers operating below 350\,MHz is most likely caused by the lack of efficient real-time data processing and search pipelines for high-time resolution data streams from wide-field radio telescopes, such as the MWA, LOFAR and LWA.

Traditional FRB searches with dish-like radio telescopes operate on a few high-time resolution timeseries (two instrumental polarisations combined into Stokes I) from the corresponding telescope beams, which is effectively small number of pixels in the sky. Such data can be comfortably processed and searched for FRBs in real-time with software pipelines typically implemented on Graphical Processing Units (GPUs) for example FREDDA (\citet{2019ascl.soft06003B}), Heimdall\footnote{\url{https://sourceforge.net/projects/heimdall-astro/}}, or \citet{2011MNRAS.417.2642M} to name a few. However, applying the same methods to low-frequency interferometers have not succeeded because of high computational requirements of forming multiple tied-array (i.e. coherent) beams tessellating the entire field of view (FoV\footnote{Here FoV is size of the field of view in 1 dimension (in units of degrees).}). 

Low-frequency radio-telescopes are also known as ``software telescopes'' because most of the signal processing is realised in software as opposed to beamforming realised by ``nature'' in dish telescopes. In order to search for FRBs or pulsars, complex voltages from individual antennas, tiles, or stations have to be beamformed in a particular direction in the sky. Such a single tied-array beam can be calculated as:

\begin{equation}
I_c(t) = \sum_{a=1}^{N_{ant}} w_c^a I_c^a(t),
\label{eq_tied_array_beam}
\end{equation}

where $I_c(t)$ is the coherent sum in channel c at time t, $I_c^a(t)$ is complex voltage from antenna a in channel c at the time t, $N_{ant}$ is the number of antennas, and  $w_c^a$ is a complex coefficient for antenna $a$ at frequency channel $c$ representing the geometrical factor to point the beam in a particular direction in the sky.
The computational cost of this operation O($N_{ant}$) scales linearly with the number of antennas (summation in equation~\ref{eq_tied_array_beam}). The resulting timeseries of complex voltages or intensities (power) can be searched for FRBs using the same software as for dish telescopes (see the earlier examples).

In order to cover the entire FoV of a low-frequency interferometer multiple tied-array beams have to be formed, and their number depends on the angular size of the tied-array beam (i.e. on the maximum baseline of the interferometer). For the ``beamformed image'' of the size $N_{px} \times N_{px}$ pixels, the computational cost is O($N_{ant} N_{px}^2$). The number of resolution elements in 1D can be calculated as $N_{px} \sim ( \text{FoV} / \delta \theta )$, where  $\delta \theta \sim \lambda/B_{max}$ is the spatial resolution of the interferometer at the observing wavelength $\lambda$ and $B_{max}$ is the maximum distance (baseline) between antennas in the interferometer. For a dish antenna $\text{FoV} \sim \lambda/D$, where $D$ is the diameter of the dish. Hence, the total number of pixels in 2D image is $\propto (B_{max}/D)^2$ and the computational cost of ``beamforming imaging'' is O($N_{ant} (B_{max}/D)^2)$). We note that for a single half-wavelength dipole (like in single SKA-Low station interferometer), $D=\lambda/2$ can be used in these considerations. 

Although, GPU-based beamforming software \citep[e.g.][]{2022PASA...39...20S} can be extremely efficient, it is still not sufficiently fast to enable real-time processing. Therefore, the presented work explores an alternative approach by forming high-time resolution time series in multiple directions using sky images obtained with standard interferometric imaging. This approach is formally nearly equivalent to forming multiple tied-array beams in the sky, and subtle mathematical differences between these two methods are outside the scope of this paper. As discussed in Section~\ref{subsec_comp_cost}, in some parts of the parameter space ($N_{ant}$, $N_{px}$ etc.) imaging can be computationally more efficient than the beamforming approach (see also Table~\ref{tab_theo_comp_cost}).  

Given that the main goal of this high-time resolution imager is to search for bright transients like FRBs, implementation of features optimising image quality and fidelity (e.g. CLEAN algorithm), which are available in general-purpose imaging packages like Common Astronomy Software Applications \citep[CASA, ][]{2022PASP..134k4501C}, MIRIAD \citep{2011ascl.soft06007S} or WSCLEAN \citep{2014MNRAS.444..606O}, is not critical. Hence, high-time resolution imager for FRB searches can be very simple and form only so called ``dirty images'', while imaging artefacts like side-lobes can be removed by subtracting a reference image of the same field (formed as a combination of previous images or prepared prior to the processing). 

Formation of high-time resolution images in real-time requires extremely efficient parallel software, which makes it well-targeted for GPUs. Nevertheless, \red{except for the} image-domain-gridding option \citep[IDG,][]{2018A&A...616A..27V} of WSCLEAN, none of the existing imagers fully utilises the compute power of modern GPUs. Although modern GPUs and associated libraries offer even an order of magnitude speed-up (see Table~\ref{tab_cpu_vs_gpu}) of Fast Fourier Transforms (FFTs) this is not fully utilised even in the IDG/GPU version of WSCLEAN. Furthermore, the existing imagers (including WSCLEAN) were not designed for high-time resolution data. Therefore, they are not suitable for FRB searches as they require input data to be converted to specific formats (such as CASA measurement sets or \texttt{UVFITS} files), which require additional input/output (I/O) operations slowing down the entire process.

The main purpose of the presented imaging software is to become a part of a streamlined GPU-based processing pipeline (Di Pietrantonio et al., in preparation), which will read input complex voltages from the archive or directly from the telescope only once and process them fully inside GPU memory in order to minimise the number of I/O operations. Finally, once the data cube of images (of multiple time steps and frequency channels) are created, dynamic spectra from all pixels (i.e time series in different directions in the sky) will be formed. Then these dynamic spectra will be searched for FRBs, pulsars or other short duration transients using one of the existing software packages or a new algorithm/software package will be developed. Apart from that, our GPU-imager performs very well (see Sections~\ref{validation_gpu_images} and \ref{sec_benchmarking}), and can also be used for other purposes where high-time resolution streams are needed.

\section{Target Instruments}
\label{sec_target_instruments}
The primary target instruments for the GPU high-time resolution imager and full pipeline are low-frequency interferometers located in the Murchison Radio-astronomy Observatory (MRO) in Western Australia (WA). In particular, the MWA and stations of the low-frequency Square Kilometre Array (SKA-Low) \citep{2009IEEEP..97.1482D}\footnote{https://www.skatelescope.org/}. Parameters of these instruments are summarised in Table~\ref{table_mwa_skalow_parameters}. Our software is publicly available at \url{https://github.com/PaCER-BLINK-Project/imager} and can be applied to data from any radio interferometer.

\subsection{SKA-Low stations}
\label{subsec_skalow_stations}

The SKA-Low telescope will comprise 512 stations, each with 256 dual polarised antennas. Since 2019, two prototype stations the Aperture Array Verification System 2 \citep[AAVS2;][]{andre_spie,2022JATIS...8a1014M}, and the Engineering Development Array 2 \citep[EDA2;][]{2022JATIS...8a1010W} have been operating and used for verification of technology, calibration procedures, sensitivity, stability testing and even early science. These stations can form all-sky images which can be used for FRB searches and lead to detections of even hundreds of FRBs per year once they are enhanced with suitable real-time search pipelines \citep{2024arXiv240104346S,2022aapr.confE...1S}. The standard data product from the stations are complex voltages in coarse ($\approx$0.94\,MHz) frequency channels. This channelisation is performed by Polyphase Filter Bank (PFB) implemented in the firmware executed in Tile Processing Units \citep[TPM;][]{2017JAI.....641014N,2017JAI.....641015C}. Long recordings of these voltages are currently impossible due to very high data rates ($\sim$9.5\,GB/s). Therefore, in order to form high-time resolution images and search for FRBs, these complex voltages have to be captured, correlated, and processed in real-time in the required time resolution.
Hence, the described GPU imager will either be applied off-line to high-time resolution visibilities saved to harddrive or in real-time as a part of a full processing pipeline. This pipeline will perform correlation and imaging, and its execution in real-time is a preferred operating mode ultimately leading to real-time FRB searches. As estimated by \citet{2024arXiv240104346S} such an all-sky FRB monitor implemented on SKA-Low stations may be able to detect even hundreds of FRBs per year. 

\subsection{The Murchison Widefield Array (MWA)}
\label{subsec_mwa}

The MWA \citep{2013PASA...30....7T,2018PASA...35...33W} is the precursor of the SKA-Low originally composed of 128 small (4$\times$4 dipoles) aperture arrays also called ``tiles''. It was recently expanded to 144 tiles with the intent of the future expansion to 256 tiles. The 16 dipoles within each tile are beamformed in analogue beamformers. The signals in X and Y polarisations are digitised and coarse channelised in receivers, then cross-correlated by the MWAX correlator \citep{2023PASA...40...19M}, which can record visibilities at time resolutions even down to 250\,ms. Besides the correlator mode, the MWA can also record coarse channelised complex voltages from individual tiles. Before commissioning of the MWAX correlator it was realised by the Voltage Capture System \citep[VCS;][]{2015PASA...32....5T}, while, presently, recording of high-time resolution voltages is implemented in the MWAX correlator itself. The MWA data archive at Pawsey Supercomputing Centre (Pawsey) contains $\sim$12\,Pb of MWA VCS from the legacy and new MWAX correlator. These data are a perfect testbed for testing the presented high-time resolution imager, and can be used to search for FRBs, pulsars or other fast transients using novel image-based approaches. For example, as estimated by \citet{2024arXiv240104346S}, the FRB search of the Southern-sky MWA Rapid Two-metre \citep[SMART;][]{2023PASA...40...21B,2023PASA...40...20B}, which can be analysed with the final high-time resolution imaging pipeline, should yield at least a few FRB detections. 

\begin{table}
\caption{Summary of parameters of the MWA and SKA-Low stations.}
\vspace{-0.3cm}
\centering
\begin{tabular}{@{}ccc@{}}
\hline
\textbf{Parameter} & \textbf{MWA} & \textbf{SKA-Low station} \\ 
\hline
Frequency range & 70 -- 300\,MHz & 50 -- 350\,MHz \\
\hline
FoV at 200 MHz & $\sim$20\degree$\times$20\degree & $\sim$12000 deg$^2$ \\
    &                                  & (full hemisphere \\
    &                                  &  at elevation $\ge$20\degree) \\
\hline
$N_{ant}$ & 128$^a$ & 256 \\
\hline
Spatial resolution $\Delta \theta$ & 1.7/1.0 $^b$ & 150 \\
at 200 MHz [arcmin]                &              &      \\

\hline
Number of pixels$^c$      &     490000 (700x700)  / & 5625 \\ 
 required to       & 1562500 (1250x1250)$^b$ & (75x75)    \\
cover FoV at 200 MHz     &                         &            \\
\hline
Number of baselines         & 16256 & 65280 \\
(without auto-correlations) &  & \\
\hline
\end{tabular}
\begin{flushleft}
\begin{tablenotes}
\small
\item $^a$ MWA tiles consist of 16 dual polarised antennas. Hence, originally it comprised total 2048 dipoles per polarisation. However, it was recently upgraded to 144 tiles with the aim of future upgrade to 256 tiles.
\item $^b$ Respectively for the MWA Phase I with maximum baseline of about 3\,km and the MWA Phase II extended configuration with maximum baseline of 5.3\,km.
\item $^c$ Assuming no oversampling (i.e. angular pixel size the same as the size of the angular size of the synthesised beam) this is proportional to $(B_{max}/D)^2$ as discussed in Section~\ref{sect_frb_search_methods}.
\end{tablenotes}
\end{flushleft}
\label{table_mwa_skalow_parameters}
\end{table}

\section{Radio-astronomy imaging}
\label{sec_radioastronomy_imaging}
The fundamentals of radio interferometry and imaging are explained in detail in one of many texts on the subject (e.g., \cite{marr2015fundamentals}, \cite{TMS}). This section provides a short summary of the most important steps of standard radio astronomy imaging which are correlation, application of various phase corrections (for cable lengths, pointing direction etc.), calibration, gridding and Fourier Transform (FT) leading to so called ``dirty images'' of the sky.  These steps are described in the context of the future GPU-based pipeline which will be based on the presented imager (Di Pietrantonio et al., in preparation). We note that there are several novel approaches to imaging, for example Efficient E-field Parallel Imaging Correlator \citep[EPIC;][]{2017MNRAS.467..715T}, which are considered in the future upgrades of the pipeline. However, the correlator code by \citet{2021A&A...656A..52R} uses tensor cores to provide an order-of-magnitude increase in processing throughput over previous GPU correlation codes. Reuse of this correlator code makes the correlation-approach computationally favourable. 

\subsection{Correlation}
\label{subsec_correlation}

In the majority of cases, the lowest level data products from modern radio telescopes are high-time resolution digitised voltages recorded by the individual antennas or tiles in the case of the MWA. These voltages can be real-valued voltages as sampled by ADCs or complex voltages resulting from FT/PFB transform of the original real-valued voltages from time to frequency domain. In the standard visibility-based imaging, these voltages are correlated and time-averaged either in real-time or off-line with modern software correlators implemented in GPUs (e.g. \citet{2013IJHPC..27..178C,2021A&A...656A..52R,2023PASA...40...19M}) or less frequently in CPUs, FPGAs or other hardware. Hardware solutions using FPGAs may be faster, but they can also be more expensive, more difficult to develop and maintain as FPGA programming expertise is generally less common and harder to develop than GPU expertise. In the most common FX correlators (F for Fourier Transform and X for cross-correlation), original real voltage samples are first Fourier Transformed (hence character \textbf{F}) to frequency domain. In the next step, channelised complex voltages from an antenna $i$ are multiplied (hence character \textbf{X}) by the corresponding (same frequency channel) complex voltages from an antenna $j$, which leads to correlation product $ij$ also known as ``visibility'' ($V_{ij}(\nu)$) calculated in $N_{ant} (N_{ant} + 1)/2$ multiplications including auto-correlations $V_{ii}(\nu)$ (correlation of voltages from an antenna $i$ with itself): 

\begin{equation}
V_{ij}(\nu) = \widehat{V}_i(\nu) \widehat{V}_j(\nu),
\label{eq_correlation}
\end{equation}

where $\widehat{V}_i(\nu)$ is FT or PFB of the original voltage samples. It can be calculated from N voltage samples using the Discrete Fourier Transform (DFT) as:

\begin{equation}
\widehat{V}_i(\nu_k) = \sum_{n=0}^{N-1} V_i(t_n) e ^{-i 2\pi \frac{k}{N} n}.
\end{equation}

In the case of the MWA and SKA-Low stations this first stage PFB is performed in the firmware of FPGAs and its computational cost is not included in the presented considerations. However, if fine channelisation into $n_{ch}$ channels is required it results in additional computational cost of FFT  O($(1/\delta t) n_{ch} log(n_{ch}))$.
Since, correlation is performed for all antenna pairs, the number of multiplications is $N_{vis} = N_{ant} (N_{ant} + 1)/2$ (including auto-correlations). Unless stated otherwise we will describe this process for a single time step. The additional time averaging of $n_t = T / \delta t$ time samples increases the computational cost by the multiplicative factor $n_t$, where T is the final integration time after averaging and $\delta t$ is the original time resolution. In the full GPU pipeline, correlation will be performed on GPU, and the number of instructions required to perform multiplications of visibilities (Equation~\ref{eq_correlation}) from a single time step is $n_{c} \sim N_{vis}/n_{core}$, where $n_{core}$ is the number of GPU cores in a specific GPU hardware (examples in Table~\ref{tab_test_gpu_devices}).

\subsection{Phase corrections and calibration}
\label{subsec:phase_corrections}

Typically, the resulting visibilities cannot be directly used for imaging, and several phase and amplitude corrections have to be applied first. In the case of the MWA  these are: (i) cable phase correction to account for different cable lengths as measured in the construction phase and stored in a configuration database, (ii) application of calibration in phase and amplitude (amplitude calibration may be skipped if correct flux density is not required) (iii) geometric phase correction, i.e. apply complex phase factor to rotate visibilities in the desired pointing direction.
For all-sky imaging with SKA-Low stations only step (ii) is required as (i) is applied in the TPM firmware and (iii) is not required for all-sky images phase-centred at zenith. These three corrections will now be described in more detail.

Firstly, a phase correction (i) needs to be applied in order to correct for different cable (or fibre) lengths between the antennas and receivers. In the case of SKA-Low stations the phase correction for different cable lengths is mostly applied in real-time in the TPM firmware as these cable lengths (or corresponding delays) can be pre-computed or measured in a standard calibration process. They remain sufficiently stable to form good quality images even without additional calibration \citep{2022JATIS...8a1010W, 2022JATIS...8a1014M,2021PASA...38...23S}. In the case of the MWA, the cable correction is currently applied post-correlation using pre-determined cable lengths provided in the metadata. These initial phase corrections (using tabulated cable lengths) are further refined by calibration (next step).

Secondly, phase and amplitude calibration (ii) is applied in order to correct for residual phase differences between the antennas and obtain correct flux scale (optional) of the final images respectively. This step typically uses calibration solutions obtained by performing dedicated calibrator observations performed close in time to the target observations so that any variations in the telescope response (e.g. due to changing ambient temperature) can be neglected. In the case of the SKA-Low stations, this step corrects for residual phase variations, for example due to temperature induced variations in electrical length of fibres with respect to the lengths applied in the firmware. The amplitude calibration provides correct flux density scale of radio sources in images; it is non-critical for FRB detection itself, but may be performed off-line to provide correctly measured flux density of the identified objects. 

Finally, the correlation as described in Section~\ref{subsec_correlation} leads to visibilities phase centred at zenith. This is sufficient to form images of the entire visible hemisphere (all-sky images) using SKA-Low stations or other aperture array with individual antennas sensitive to nearly entire sky. However, for instruments like the MWA with a smaller FoV, a geometric phase correction (iii) has 
to be applied to visibilities (post-correlation) in order to rotate the phase centre to the centre of the primary beam where the telescope was pointing (as set by the settings of MWA analogue beamforming). 

Only after all these corrections are applied, the visibilities are ready for the imaging step. It is worth noting that antenna-level phase/amplitude corrections (i.e. (i) and (ii)) can be applied to visibilities (post-correlation) or voltages (pre-correlation), which may be computationally more efficient. In the presented software, most of these corrections are already implemented as GPU kernels and are executed post-correlation. However, we will consider moving (i) and/or (ii) into the pre-correlation stage (i.e. apply to voltages) in order to further optimise the code.

\begin{table*}
\caption{Different GPU architectures used for testing and benchmarking of the presented imager.} 
\begin{tabular}{@{}ccccc@{}}
\hline
\textbf{System or }  & \textbf{GPU model} & \textbf{\# GPU / }   & \textbf{CUDA / HIP} & \textbf{Memory} \\
\textbf{HPC name}    &                    & \textbf{Tensor Cores$^a$} & \textbf{Version} & \textbf{Bandwidth} \\
                     &                    &                           &                  & \textbf{[GB/s]} \\   
\hline
Setonix$^b$ & AMD Instinct MI250X GPUs$^c$ & 14080 / 880 & ROCM 5.4.3 & 3200 \\ 
Garrawarla$^d$ & NVIDIA Tesla V100 32GB GPU & 5120 / 640 & CUDA 10.1 & 1134 \\ 
Workstation & NVIDIA GeForce RTX 2060 & 2176 / 272 & CUDA 10.1 & 336 \\
Laptop A & NVIDIA GeForce RTX 3070  & 5888 / 0     & CUDA 11.6 & 448 \\
Laptop B & NVIDIA GeForce GTX 1060  & 1280 / 0     & CUDA  9.1.85 & 192 \\
\hline
\end{tabular}
\begin{flushleft}
\begin{tablenotes}
\small
\item $^a$ NVIDIA Tensor Core are called Matrix Cores in AMD nomenclature. Similarly shader cores are AMD counterparts of CUDA codes. Thus, here a GPU core was used as a general term.
\item $^b$ \url{https://pawsey.org.au/systems/setonix/}
\item $^c$ \url{https://www.amd.com/content/dam/amd/en/documents/instinct-business-docs/white-papers/amd-cdna2-white-paper.pdf}
\item $^d$ \url{https://pawsey.org.au/systems/garrawarla/}
\end{tablenotes}
\end{flushleft}
\label{tab_test_gpu_devices}
\end{table*}

\subsection{Imaging}
\label{subsec:imaging}

This section provides a short summary of the imaging steps for the case of a single frequency channel (monochromatic wave) and single time step. It can be shown (e.g., \cite{marr2015fundamentals}, \cite{TMS}) that the visibilities and sky brightness form a Fourier pair. This can be expressed as van Cittert-Zernike theorem, which in its simplified form can be written as: 

\begin{equation}
  V(u,v,w)  = \iint A(l,m) I(l,m) e^{-i2\pi(ul + vm + wn)} \frac{\,dl\,dm}{n},
\label{eq_van_cittert}
\end{equation}

where $(u,v)$ are baseline coordinates expressed in units of wavelengths, present in a right-handed coordinate system with z-axis pointing towards the observed source (phase centre), v is measured toward the north in the plane defined by the origin, source and the pole, and u is determined by axes w and v, $V(u,v,w)$ is the visibility as a function of (u,v,w) coordinates, $(l,m,n)$ are directional cosines, measured with respect to axes u, v and w respectively. $I(l,m)$ is the sky brightness distribution corresponding to an image of the sky. The third directional cosine $n$ can be expressed in terms of the other two ($n=\sqrt{1 - l^2 - m^2}$), and for small FoV $n \approx 1$. For a co-planar array and an all-sky image phase centred at zenith $w \approx 0$, which is the case of all-sky imaging with SKA-Low stations, and in this case equation ~\ref{eq_van_cittert} can be simplified to:

\begin{equation}
  V(u,v,0)  = \iint \frac{A(l,m) I(l,m)}{\sqrt{1 - l^2 - m^2}} e^{-i2\pi(ul + vm)} \,dl\,dm,
\label{eq_van_cittert2}
\end{equation}

where approximation $e^{-i2\pi w} \approx 1$ was applied for $w \approx 0$.
This equation shows that the visibility function $V(u,v)$ is a Fourier Transform of the function $I^{^{\prime}}(l,m) = A(l,m) I(l,m)/n$. Therefore, the function $I^{^{\prime}}(l,m)$ can be calculated as an inverse 2D Fourier Transform of the visibility function $V(u,v)$. The resulting image of the sky is called ``dirty image'' because in practise $V(u,v)$ is not measured at every point on the UV plane, but only sampled at multiple (u,v) points corresponding to existing pairs of antennas (baselines) in the specific interferometer. Hence, the measured visibility function $V_m(u,v) = S(u,v) V(u,v)$, where $S(u,v)$ is the sampling function equal to 1 at (u,v) points where the baseline exists and zero otherwise. As a result, the inverse FT of  $V_m(u,v)$ is:

\begin{equation}
  FT^{-1}\left( V_m(u,v) \right)= I^{^{\prime}}(l,m) * FT^{-1}\left( S(u,v) \right),
\label{eq_dirty_image}
\end{equation}

where $*$ is convolution. Therefore, a simple 2D FT of the measured visibilities equals $I^{^{\prime}}(l,m)$ convolved with the inverse Fourier Transform of $S(u,v)$, the so called ``dirty beam''. In order to remove side-lobes and other artefacts and recover the function $I^{^{\prime}}(l,m)$ (and later $I(l,m)$) non-linear de-convolution algorithms, such as CLEAN~\citep{1974A&AS...15..417H} have to be applied. These algorithms are very computationally expensive and, therefore, not implemented in the presented high-time resolution imager. Fortunately, the presented imager will be applied to transient searches and can take advantage of the fact that artefacts present in ``dirty images'' can be removed by subtracting a reference image. \red{Such a reference image can be the preceding ``dirty image'' image, some form of an average of the sequence of previous images recorded during the same observation or a model image. However, in order to reproduce the same artefacts a model image would have to be generated for the same array configuration and with the same imaging parameters (no CLEANing etc.). Hence, a reference image obtained from the very same data is likely to be the most practical option.} Alternatively, reference or model \red{(subject to earlier mentioned limitations)} visibilities can be subtracted before applying inverse an FT. Therefore, for the presented transient science applications de-convolution is not strictly required. Similarly, transient/FRB searches can be performed on non beam-corrected $I^{^{\prime}}(l,m)$ sky images, and only once FRB candidate is detected beam correction (division by A(l,m)) can be applied in order to measure correct flux density of the detected objects.  

In practice, equation~\ref{eq_van_cittert} has to be calculated numerically, and the Fast Fourier Transform (FFT) algorithm \citep{Cooley1965AnAF} is the most efficient way to do this. Its computational complexity is $O(NM log(NM))$, where N and M are the dimensions of the UV grid. An FFT requires the input data (i.e. complex visibilities) to be placed on a regularly spaced grid in the UV-plane in the process called gridding (Section~\ref{subsec_gridding}). 


\subsection{Gridding}
\label{subsec_gridding}

Before 2D FFT can be performed, complex visibilities have to be placed in a regularly spaced cells on UV grid in the process called gridding, which is summarised in this section.

\subsubsection{Gridding Parameters}
\label{subsec_gridding_in_depth}

The output sky images are typically $N_{px} \times N_{px}$ square arrays of pixels, where each pixel has angular size $\Delta$x $\times$ $\Delta$x. Thus, the angular size of the entire sky image is $(N_{px}\Delta x)^2$, while the angular size of the synthesised beam is $\Delta \theta = \lambda / B_{max} = 1 / u_{max}$, where $\lambda$ is the observing wavelength and $B_{max}$ is the maximum baseline. The longest baselines ($B_{max}$) correspond to maximum angular resolution (smallest $\Delta \theta$). 

In order to Nyquist sample the longest baselines (i.e $u_{max}$ and $v_{max}$) the FWHM of the synthesised beam has to be over-sampled by at least a factor of two. Hence, the condition for the pixel size is:

\begin{equation}
  \Delta x  \le \Delta \theta / 2,
\end{equation}

which after substituting $\Delta \theta = 1 / u_{max}$ can be written as:

\begin{equation}
  \Delta x  \le \frac{1}{2u_{max}}.
\end{equation}

Typically, the synthesised beam is over-sampled by a factor between 3 or 5, which can be realised by specifying parameters of the imager. The dimensions of the UV-grid cells are determined by the maximum angular dimensions of the sky image given by: 
\begin{equation}\label{delta_u}
  \Delta u  = \Delta v = \frac{1}{N_{px}\Delta x}.
\end{equation}

In the case of limited FoV ($\sim 25$\degree$\times25$\degree at 150\,MHz) images from the MWA, Fourier Transform of sampled visibility function will lead to aliasing effects where sources from outside the FoV are aliased into the final sky image \citep{1984AJ.....89.1076S,1984iimp.conf..333S}. In order to mitigate these effects gridded visibilities are usually convolved with a gridding kernel, which increases computational cost of gridding by a factor $N_{kern}$ (Table~\ref{tab_theo_comp_cost}). This is not required in imaging of the entire visible hemisphere (i.e. all-sky imaging), because there are no sources outside the FoV which could be aliased into the FoV. This significantly simplifies the procedure of forming all-sky images with the SKA-Low stations.

\subsubsection{Visibility Weighting}
\label{weights}
Visibilities are gridded such that each cell in the UV-grid satisfies one of the three conditions: (i) if there are no visibilities corresponding to that cell the cell will have a zero value, (ii) if there is one visibility, the cell will have that value, or (iii) if there are multiple visibilities assigned to that cell, the value in this cell will be a weighted sum of these visibilities. The choice of the weighting scheme can be specified by parameters of the imager, and currently natural and uniform weighting schemes have been implemented. In natural weighting visibilities assigned to specific cell are summed, and they contribute with the same weights. Hence, this weighting uses all the available information which minimises system noise, but leads to lower spatial resolution due to larger contribution from, more common, shorter baselines. 

On the other hand in the uniform weighting, visibilites are weighted by the UV area. Thus, all the baselines contribute with equal weights. This improves the spatial resolution (contribution from longer baselines is effectively ``up-weighted''), but may lead to slightly higher system noise. It is worth noting that the impact of uniform weighting on final image noise is a combination of higher system noise and reduced confusion noise due to better spatial resolution. Since the imager is intended to be simple and applied to FRB and transient searches, we have not implemented other weighting schemes.

\begin{figure}
\centering
\includegraphics[width=\textwidth]{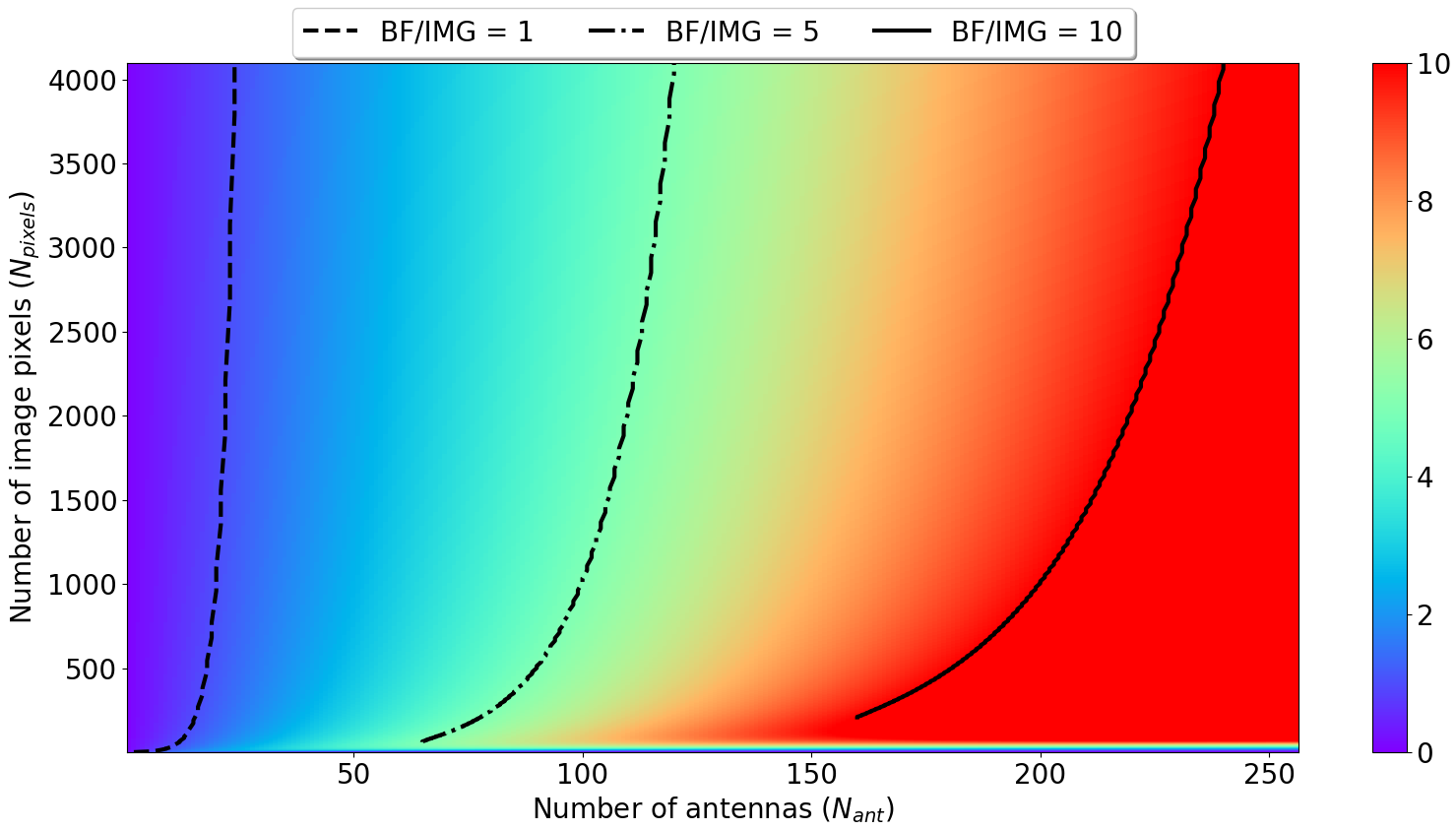}
\caption{Ratio of theoretical compute cost of beamforming (label BF) to imaging (label IMG) as a function of number of antennas ($N_{ant}$) and number of pixels in 1D ($N_{px}$) based on the total cost equations in the Table~\ref{tab_theo_comp_cost}. The black curves superimposed on the graph indicate where the ratio of theoretical cost of beamforming and imaging is of the order of one (dashed line), 5 (dashed-dotted line) and 10 (solid line). The plot shows that for 128 and 256 antennas (MWA and SKA-Low station respectively) the number of operations in beamforming is respectively at least around 5 and 10 times larger than in imaging indicating that standard visibility-based imaging should be more efficient than beamforming in these regions of parameter space.}
\label{fig_beamforming_vs_imaging}
\end{figure}

\subsubsection{Theoretical Computational Cost}
\label{subsec_comp_cost}

This section summarises theoretical computational costs of the main steps of the imaging described in the previous section. These theoretical costs are also compared for different numbers of antennas and pixels (Table~\ref{tab_theo_comp_cost} and Figure~\ref{fig_beamforming_vs_imaging}). Figure~\ref{fig_beamforming_vs_imaging} shows the ratio of theoretical computational costs of beamforming to visibility based imaging as a function of number of antennas and image size. It is clear that for 128 and 256 antennas (MWA and SKA-Low station respectively) the number of operations in beamforming is respectively at least around 5 and 10 times larger than in imaging indicating that standard visibility-based imaging should be more efficient than beamforming in these regions of parameter space.

\begin{table*}
\caption{The summary of the theoretical costs of the main steps of imaging and beamforming. $N_{ant}$ is the number of antennas (or MWA tiles), $N_{px}$ is 1D dimensional number of resolution elements (pixels). Hence, for square images the total number of pixels is $N_{px}^2$. $N_{kern}$ is the size (total number of pixels) of the convolving kernel. It can be seen that computational cost of correlation and gridding is independent of the image size. It depends only on the number of visibilities to be gridded, which equals number of baselines $N_b$ directly related to the number of antennas as $N_b = \sfrac{1}{2} N_{ant} (N_{ant} - 1)$ (excluding auto-corrrelations). The total computational cost is dominated by the FFT, and is directly related to the total number of pixels in the final sky images. For the presented image sizes, the imaging requires a few ($\sim$ 5 -- 8) times less operations than beamforming for 128 antennas (MWA) and even order of magnitude less ($\sim$ 11 -- 15 times) for 256 antennas (SKA-Low stations). This is because the total cost of both is dominated by the component $\sim \alpha N_{px}^2$, where $\alpha=N_{ant}$ for beamforming and $\alpha=log(N_{px}^2)$ for imaging ($log$ is the logarithm to the base 2). Hence, for a given image size $N_{px}^2$ beamforming dominates when $N_{ant} \ge log(N_{px}^2)$. For example, for image size 180x180 beamforming dominates when $N_{ant} \gtrsim$15.} 
\begin{tabular}{@{}cccccc@{}}
\hline\hline
\textbf{Step}        & \textbf{Time} & \textbf{N$_{a}$} & \textbf{Number of}  & \textbf{Number of} & \textbf{Number of} \\
                     & \textbf{Complexity}        &                  & \textbf{Operations} & \textbf{Operations}   & \textbf{Operations} \\
                     &                      &                  & for \textbf{$N_{px}=$180} & for \textbf{$N_{px}=$1024} & for \textbf{$N_{px}=$4096}  \\
                     & & & ($N_{px}^2 = 32400$) & ($N_{px}^2 = 1048576$) & ($N_{px}^2 = 16777216$) \\
\hline
Correlation  & O$(\sfrac{1}{2} N_{ant} (N_{ant} - 1))$ & 128 & 8128 & 8128 & 8128 \\
($n_{corr}$) &                                     & 256 & 32640 & 32640 & 32640 \\
\hline
Gridding     & O$(\sfrac{1}{2} N_{ant} (N_{ant} - 1) N_{kern})$ & 128 & 8128 & 8128 & 8128 \\
($n_{grid}$) &                                              & 256 & 32640 & 32640 & 32640 \\
\hline
FFT          & O$(N_{px}^{2} log_{2}(N_{px}^{2}))$ & 128 & 485472 &  20971520 & 402653184 \\
($n_{fft}$) &                                     & 256 & 485472 & 20971520 & 402653184 \\
\hline
\textbf{Imaging Total}       &   O($n_{corr} + n_{grid} + n_{fft}$) & 128 & 501728 & 20987776 & 402669440 \\        
($n_{tot}$) &                                     & 256 & 550752 & 21036800 & 402718464 \\                    
\hline
\hline
\textbf{Beamforming Total} &   O$(N_{px}^{2} N_{ant})$               & 128 & 4147200 & 134217728 & 2147483648 \\   
 ($n_{bf}$) &                                     & 256 & 8294400 & 268435456 & 4294967296 \\     
\hline
\hline
\textbf{Ratio}       &   O($N_{ant} / log_{2}(N_{px}^{2}))$    & 128 & 8.3 & 6.4 & 5.3 \\
($n_{bf} / n_{tot}$) &                            & 256 & 15.1 & 12.8 & 10.7 \\
\hline
\hline
\end{tabular}
\label{tab_theo_comp_cost}
\end{table*}

However, the real cost of the beamforming and imaging operations depends on the actual implementation. 
For example, modern GPUs such as NVIDIA V100, with up to 5120 cores allow performing correlation of voltages from 128 (the MWA) and 256 antennas (EDA2) in O($N_{ant}^2/N_{cores}$) number of instructions, which is O(1) for both MWA and EDA2. Hence, correlation on GPU can be performed faster than the mathematical cost of sequential operation, and is mainly limited by the memory bandwidth of the GPUs.

\section{Implementation of the imager}
\label{sec_implementation}

The initial version of the imager was implemented entirely on CPU in order to test and validate the code on real and simulated data. Once it was tested and validated, the main imaging steps were ported to GPU. Both versions were developed in C++. This section describes the CPU and GPU versions, and presents the tests and validations performed on real and simulated data from the MWA and SKA-Low prototype station EDA2.

\begin{figure*}
\centering
\includegraphics[width=\textwidth]{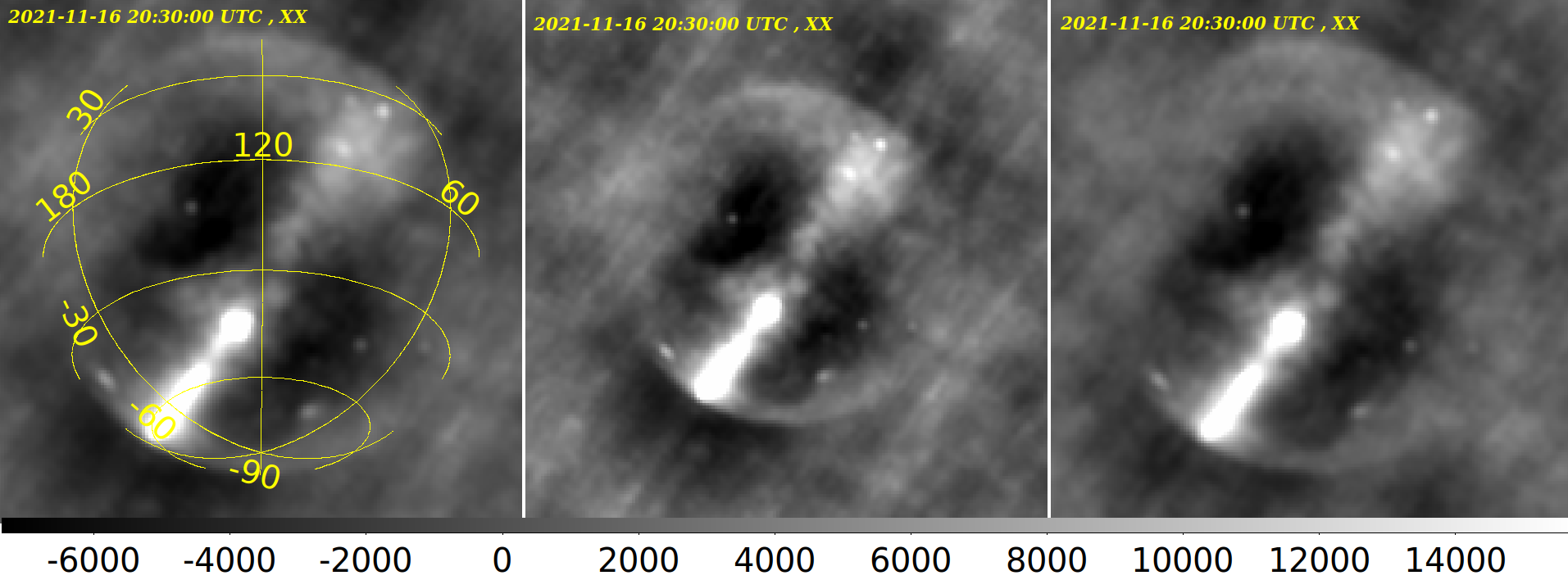}
\caption{A comparison of the all-sky images from EDA2 visibilities at 160\,MHz recorded on 2021-11-16 20:30 UTC. \textit{Left}: Image produced with MIRIAD. \textit{Centre}: Image produced with CASA. \textit{Right}: Image produced with BLINK imager presented in this paper. All are 180x180 pixels dirty images in natural weighting. \red{The images are not expected to be the same because MIRIAD and CASA apply a gridding kernel, which has not been implemented in the BLINK imager yet (as discussed in Section~\ref{subsec_cpu_imager}). Therefore, the differences between the BLINK and MIRIAD/CASA images are of the order of 10 -- 20\%.}}
\label{fig_eda2_real_data_blink_casa_wsclean}
\end{figure*}

\subsection{CPU Imager}
\label{subsec_cpu_imager}

The first version of the imager was designed to create images of the entire visible hemisphere (all-sky images) using visibilities from SKA-Low stations (mostly EDA2). \red{The main reason for starting with SKA-Low stations data} was the simplicity of all-sky imaging, which is mathematically correct without small FoV approximation, and does not require additional convolution kernel in gridding (no aliasing of out-of-FoV sources in all-sky images). Additionally, it was decided early on to implement \red{the CPU version} first, validate it on real and simulated data, develop reference datasets and expected template output data (sky images, gridded visibilities etc.). Based on these datasets, test cases were created and used in the development process, including the validation of the GPU version. These test cases were included in the build process, and once the CPU version was tested and validated, the GPU version was developed and tested using the same test cases with reference datasets. Any variations in the results were carefully investigated and led to identification of ``bugs'' or useful insights into inherent differences in GPU processing with respect to CPU.  

The CPU version was implemented as a single threaded application, and the main components of the imaging process (the gridding and FFT) were realised in CPU. The FFT was implemented using functions \texttt{fftw\_plan\_many\_dft} and \texttt{fftw\_execute\_dft} from the \texttt{fftw}\endnote{\href{https://www.fftw.org/}{https://www.fftw.org/}} library. Initially, the software could only generate all-sky images for a single time stamp and frequency channel. However, it was later expanded to process EDA2 visibilities in multiple fine channels and timesteps, and was used to process a few hours of data, which will be described in the future publication (Sokolowski et al., in preparation). In the next step the imaging code was generalised to non all-sky cases and tested on real and simulated MWA data.

\begin{figure*}[t]
\begin{center}
\includegraphics[width=0.95\textwidth,angle=0]{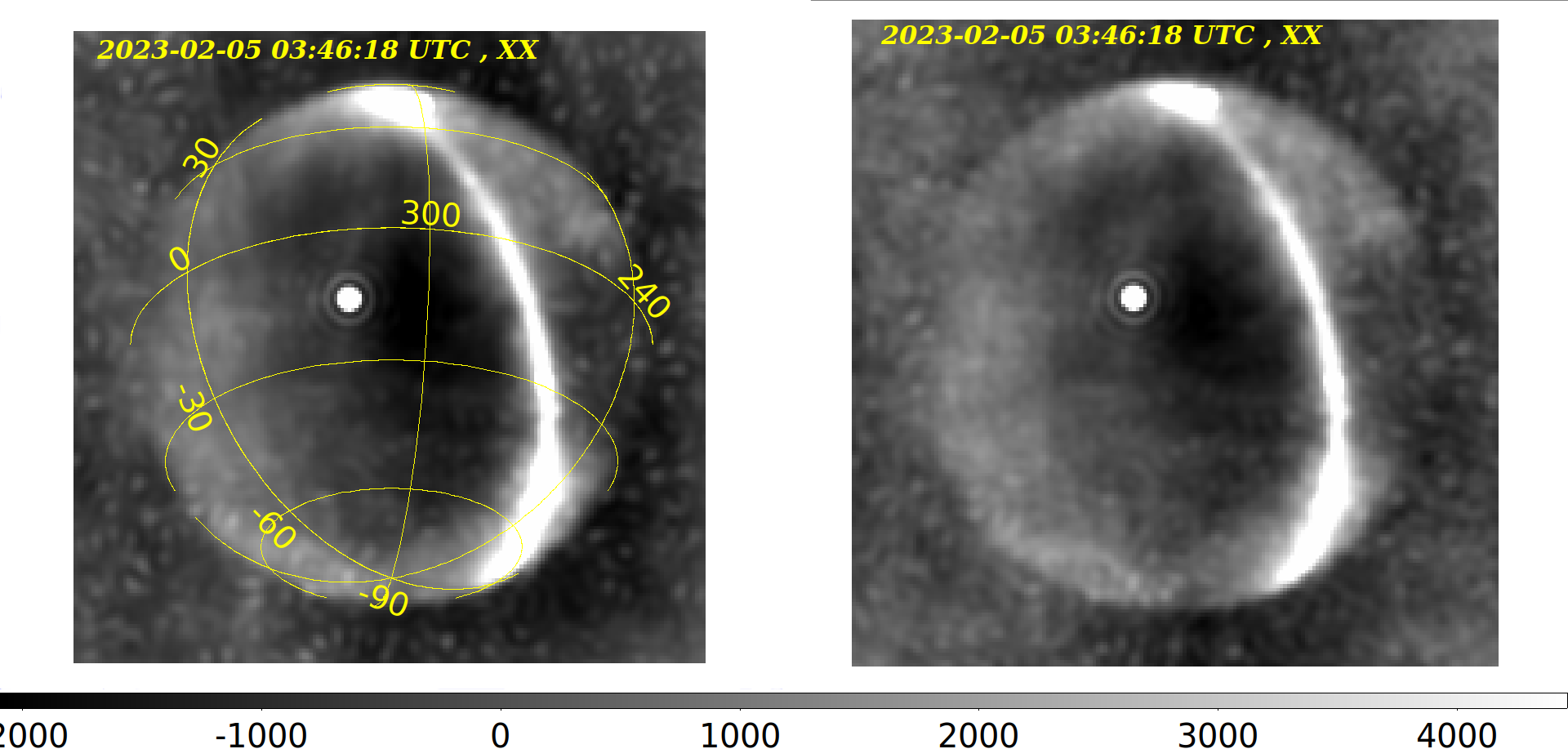} 
\caption{Sky images in X polarisation produced from EDA2 visibilities at 160\,MHz generated in MIRIAD and using frequency-scaled ``HASLAM map'' as a sky model. The image corresponds to real EDA2 data recorded on 2023-02-05 03:46:18 UTC. \textit{Left}: Image generated with MIRIAD task invert. \textit{Right}: Image generated with the CPU version of the BLINK imager presented in this paper.}
\label{fig_eda2_simulated_data}
\end{center}
\end{figure*}

\subsubsection{Validation of CPU imager on real and simulated data}
\label{sec_cpu_imager_validation}

The initial validation was performed using real data from EDA2, and the images from the presented BLINK imager were compared to well established radio-astronomy imagers (CASA and MIRIAD) applied to the same data and using the same imaging settings (i.e. ``dirty image'' and natural weighting).
The comparison of the example validation images is shown in Figure~\ref{fig_eda2_real_data_blink_casa_wsclean}.

In the next steps, EDA2 visibilities were simulated using MIRIAD task \texttt{uvmodel} and an all-sky model sky image was generated using the all-sky map at 408\,MHz \citep[][the so called ``HASLAM map'']{1982A&AS...47....1H} scaled down to low frequencies using a spectral index of $-2.55$ \cite{spectral_indexASU2019}.
The generated visibilities were imaged with both MIRIAD and BLINK imagers and their comparison is shown in Figure~\ref{fig_eda2_simulated_data}.

\begin{figure*}[t]
\begin{center}
\includegraphics[width=0.95\textwidth,angle=0]{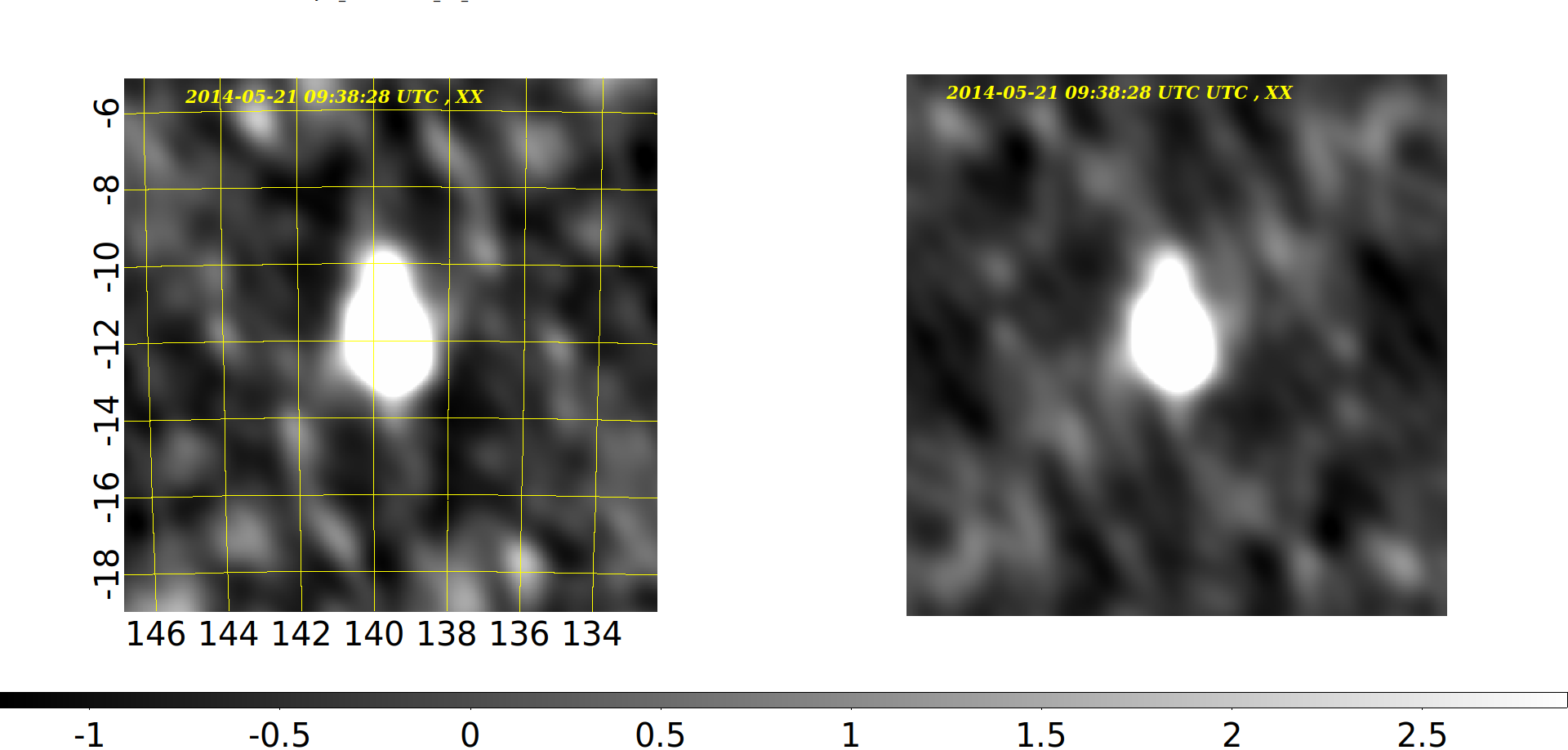} 
\caption{Sky images in X polarisation generated from MWA visibilities at 154\,MHz simulated in CASA based on the model image of Hydra-A radio galaxy. \textit{Left}: Image generated with WSCLEAN. \textit{Right}: Image generated with the CPU version of the BLINK imager presented in this paper.}
\label{fig_mwa_simulated_data}
\end{center}
\end{figure*}

Similar verifications were performed on the MWA visibilities simulated with CASA tasks \texttt{simobserve} using the model image of Hydra-A radio galaxy. The resulting images were imaged with CASA task \texttt{simanalyze}, WSCLEAN and BLINK imagers (all dirty images in natural weighting), and their comparison is shown in Figure~\ref{fig_mwa_simulated_data}. 

All the above tests gave us confidence that the images formed by the BLINK imager are correct, and the test data were used to develop test cases, which were included into the software build procedure (as a part of \texttt{CMake} process). This turned out to be extremely useful feature, which sped-up and made the software development process more robust. In summary, after the software is build (on any system) the imager is executed on a few test datasets and output data (i.e. sky images) are compared to the template images, which are part of the test dataset. If the final sky images are the same to within small limits the test passes, and otherwise the test fails, which indicates that some ``bug'' was introduced in the newest (currently compiled) version of the code. This allows to discover errors (``bugs'') in the code immediately after introducing them (assuming the code is compiled after small incremental changes), and potentially correct them straightaway using the knowledge of which particular parts of the code were modified.

\subsection{GPU imager}
\label{sec_gpu_imager}

\begin{figure}[t]
\begin{center}
\includegraphics[width=0.95\textwidth,angle=0]{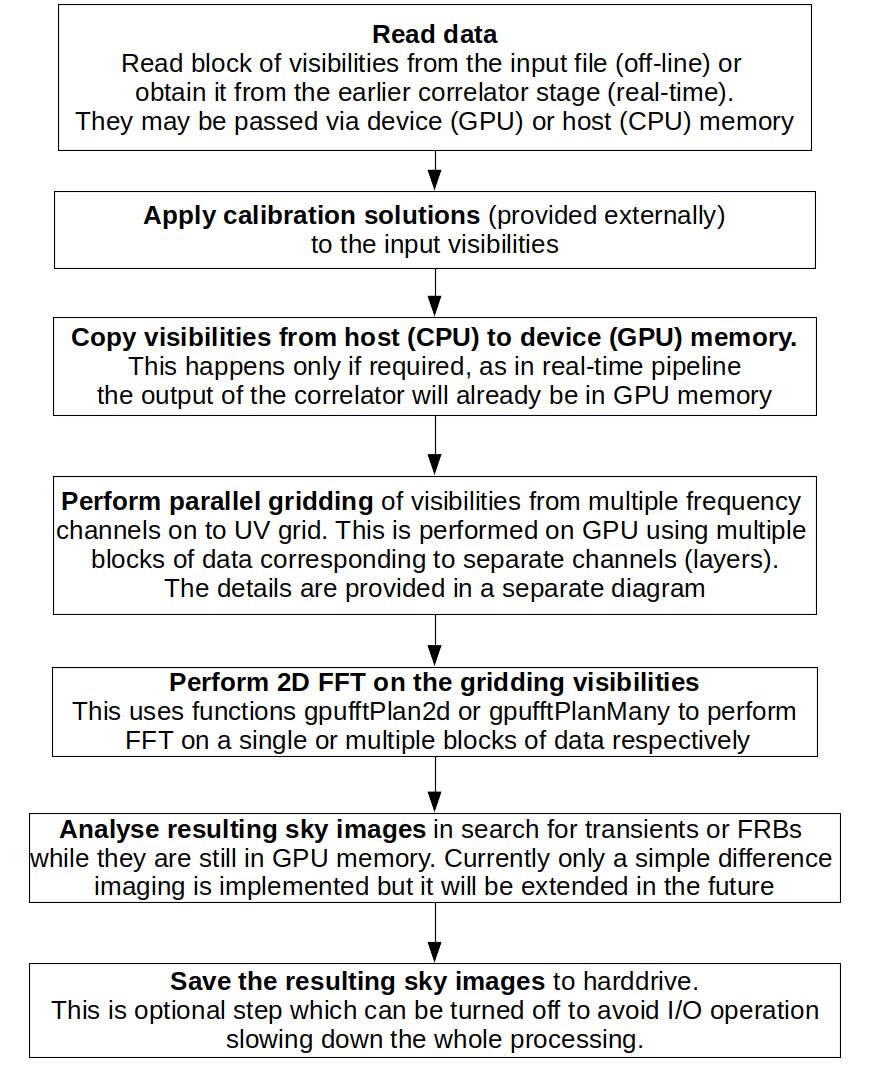} 
\caption{Block diagram of the main steps in the GPU version of the imager for multiple frequency channels.}
\label{fig_gpu_imager_diagram}
\end{center}
\end{figure}

The first version of the GPU imager was implemented on the basis of the CPU imager by gradually implementing the main steps in GPU using Computer Unified Device Architecture (CUDA)\endnote{\href{http://nvidia.com/cuda./}{http://nvidia.com/cuda}} programming environment, where the code in CUDA is written in C++. The conversion process started with the calls to \texttt{fftw} library, which were replaced with the corresponding calls to functions from the \texttt{cuFFT} library\footnote{\url{https://developer.nvidia.com/cufft}}. In the next step, sequential gridding function for CPU was implemented as a GPU kernel in CUDA.  During this process the test cases developed for the CPU imager were used to validate the results of implementation as each part of the code was ported to GPU. The block diagram of the GPU version of the imager is shown in Figure~\ref{fig_gpu_imager_diagram}.

The code was later translated to AMD programming framework Heterogeneous-Compute Interface for Portability (HIP)\endnote{\href{https://github.com/ROCm/HIP}{https://github.com/ROCm/HIP}}, and a separate branch was created. Hence, the imager could be executed on NVIDIA and AMD hardware using respectively CUDA and HIP programming frameworks. To unify the CUDA and HIP codebases a set of \texttt{C++} macros named \texttt{gpu<FunctionName>} was implemented and used in the code to replace specific CUDA and HIP calls. An \texttt{\#ifdef} statement in the include file \texttt{gpu\_macros.hpp} selects a set of CUDA or HIP versions of the function depending if the compiler is \texttt{nvcc} or \texttt{hipcc} respectively. 
For example, the calls to functions \texttt{cudaMemcpy} and \texttt{hipMemcpy} in the CUDA and HIP branches respectively, were replaced by \texttt{gpuMemcpy} in the final merged branch supporting both frameworks. Therefore, later in the text, names of functions starting with \texttt{gpu<FunctionName>} should be interpreted as expanding to \texttt{cuda<FunctionName>} or \texttt{hip<FunctionName>} in the CUDA or HIP framework/compiler respectively. The imager can be compiled and executed in one of the two frameworks depending on the available hardware and software. We note that the HIP framework also supports using the NVIDIA CUDA as a back-end, but it has to be installed on a target system, which is not always the case. Therefore, the macros make our package much more flexible as it can be compiled for both NVIDIA and AMD GPUs regardless if HIP framework is installed on the target system or not. 

The code was originally developed in a notebook/desktop environment with NVIDIA GPU, and later it was also compiled, validated, and benchmarked on the now-retired Topaz, Garrawarla\footnote{\url{https://pawsey.org.au/systems/garrawarla/}} and Setonix\footnote{\url{https://pawsey.org.au/systems/setonix/}} supercomputers at Pawsey. The list of the test systems is provided in Table~\ref{tab_test_gpu_devices}. 

The following subsections describe implementations of the main parts of imaging in the GPU version.

\subsubsection{GPU Gridding Kernel}
\label{sec_gpu_gridding_kernel}

Gridding places complex visibility values on the UV grid in the cell with coordinates $(u,v)$, which are coordinates of a baseline vector, i.e. difference between the antenna positions expressed in wavelengths. The main steps in the GPU gridding kernel\footnote{Kernels are functions executed by a GPU.} are: 

\begin{enumerate}
    \item Calculate coordinates $(x,y)$ of the specific baseline $(u,v)$ on the UV grid. The correlation matrix is a 2D array of visibilities indexed by the first and second antenna, whose signals were correlated to obtain the visibility that is about to be added to a UV cell (``gridded''). Due to the requirements for the formatting of the FFT input (both in FFTW and cuFFT/hipFFT versions) the visibilities have to be gridded in such a way that the center bin corresponds to zero spatial frequency (so-called DC term). Hence, in order to avoid moving the data after gridded, the $(x,y)$ indexes are calculated to satisfy these requirements. \\
    
    \item Add visibility value to the specific UV cell $(x,y)$ in the 2D complex array (UV grid). This UV array is initialised with zeros and the additions are performed with \texttt{atomicAdd} to ensure that two threads do not modify the same cell of the array in the same time. \\

    \item Increment the counter of the number of visibilities added to the specific UV cell, which can be used later to apply selected weighting schema (not required in the default natural weighting).
\end{enumerate}

Once the gridding kernel has completed, the selected weighting can be applied (for weightings other than natural) and the gridded visibilities are ready for the FFT step.

On a GPU, each visibility value can be gridded by a dedicated thread, each running on a separate core and executing the same kernel instruction in parallel. The number of available GPU cores ($N_{core}$) depends on specific device (summary in Table~\ref{tab_test_gpu_devices}), while the number of visibility points depends on the number of antennas ($N_{vis} = \sfrac{1}{2} N_{ant} (N_{ant} - 1)$ when auto-correlations are excluded). Typically, $N_{vis} > N_{core}$, and it is not possible to grid all visibility points having all cores simultaneously execute the same operation for different baselines. This is a standard scenario as for most of the problems the number of data points (e.g. size of the input array) is larger than the maximum number of threads which can be executed simultaneously. Therefore, GPU threads are grouped into \texttt{nBlocks} blocks of \texttt{NTHREADS} threads (typically $1024$), and the threads within a single block of threads are executed simultaneously while blocks of threads are executed one by one (sequentially). Hence, in order to grid $N_{vis}$ visibility values, there are
\begin{center}
\ttfamily 
nBlocks = ($N_{vis}$ + NTHREADS -1)/NTHREADS
\end{center}

blocks of threads, which ensures that the total number of threads across all blocks is greater than or equal to $N_{vis}$. Usually not all threads in the last block are mapped to a visibility to be gridded. For this reason, in the GPU kernel there exists a safeguard conditional statement to avoid out-of-bounds memory access when a thread (global) index is greater than the maximum visibility index in the input array.

Indexes of a GPU block and thread are available inside the GPU kernel and are used to address corresponding data (visibility points in this case) to be processed\footnote{Refer to \url{https://docs.nvidia.com/cuda/cuda-c-programming-guide/index.html} for more information about the main principles of GPU programming.}. The maximum number of GPU threads that can be executed concurrently depends on several factors, such as the total number of GPU cores in a specific device (Table~\ref{tab_test_gpu_devices}), register and memory resources required by the GPU kernel etc. In the simplest case where different threads write to different cells of the output array, no synchronisation mechanisms are required. However, in some cases  simultaneously running threads may need to write to the same output variables or array cells. Hence, memory access synchronisation mechanisms must be adopted to avoid unpredictable results. This is in fact the case of the gridding kernel, where each thread processes specific cell of the correlation matrix, and in some cases different GPU threads may try to write to the same UV cell as pairs of different antennas may have baselines with nearly the same (u,v) coordinates.
Synchronisation is achieved using the \texttt{atomicAdd} function (present in both CUDA and HIP). It should be noted that this function may slow down the code, but currently this is not the main bottleneck. We plan to replace it in the future with the more efficient parallel reduction algorithm \citep{HWU2023211}, which ensures that data (visibilities in this case) are not written to the same memory cells in the same time.

\subsubsection{Fourier Transform of gridded visibilities}
\label{sec_fft_of_gridded_vis}

As described in Section~\ref{sec_radioastronomy_imaging}, after visibilities are gridded on a regular grid, an FFT can be applied to form a dirty image. In the GPU imager the FFT plan is created using functions \texttt{gpufftPlan} (expanding to cudafftPlan2d or hipfftPlan2d) or \texttt{gpufftPlanMany} (expanding to cudafftPlanMany or hipfftPlanMany) for a single and multiple images respectively. These functions correspond to \texttt{fftw\_plan\_dft\_2d} and \texttt{fftw\_plan\_many\_dft} in the CPU implementation. Plan creation is performed once (at the first execution of FFT), therefore its computational cost can be neglected. The actual execution of FFT is performed by the function \texttt{gpufftExecC2C} in the GPU version, which corresponds to \texttt{fftw\_execute} in the CPU version.

The real part of the output of the FFT is a sky image. However, similarly to the input stage, the resulting output array has to be re-organised by the so called FFT shift operation in order to move the pixels so that the center of the image corresponds to the phase centre (by default it is in the corner of the image). This operation is required at least for the validations and comparisons of the resulting sky images with the template test images and output from other software packages. However, in order to avoid additional computational cost (O($N_{px}^2$)), transient search operations can also be performed on the direct output from the FFT. Only images in which interesting candidates were identified would be ``FFT-shifted'' for further inspection and analysis.

\begin{figure*}
\begin{center}
\includegraphics[width=0.95\textwidth,angle=0]{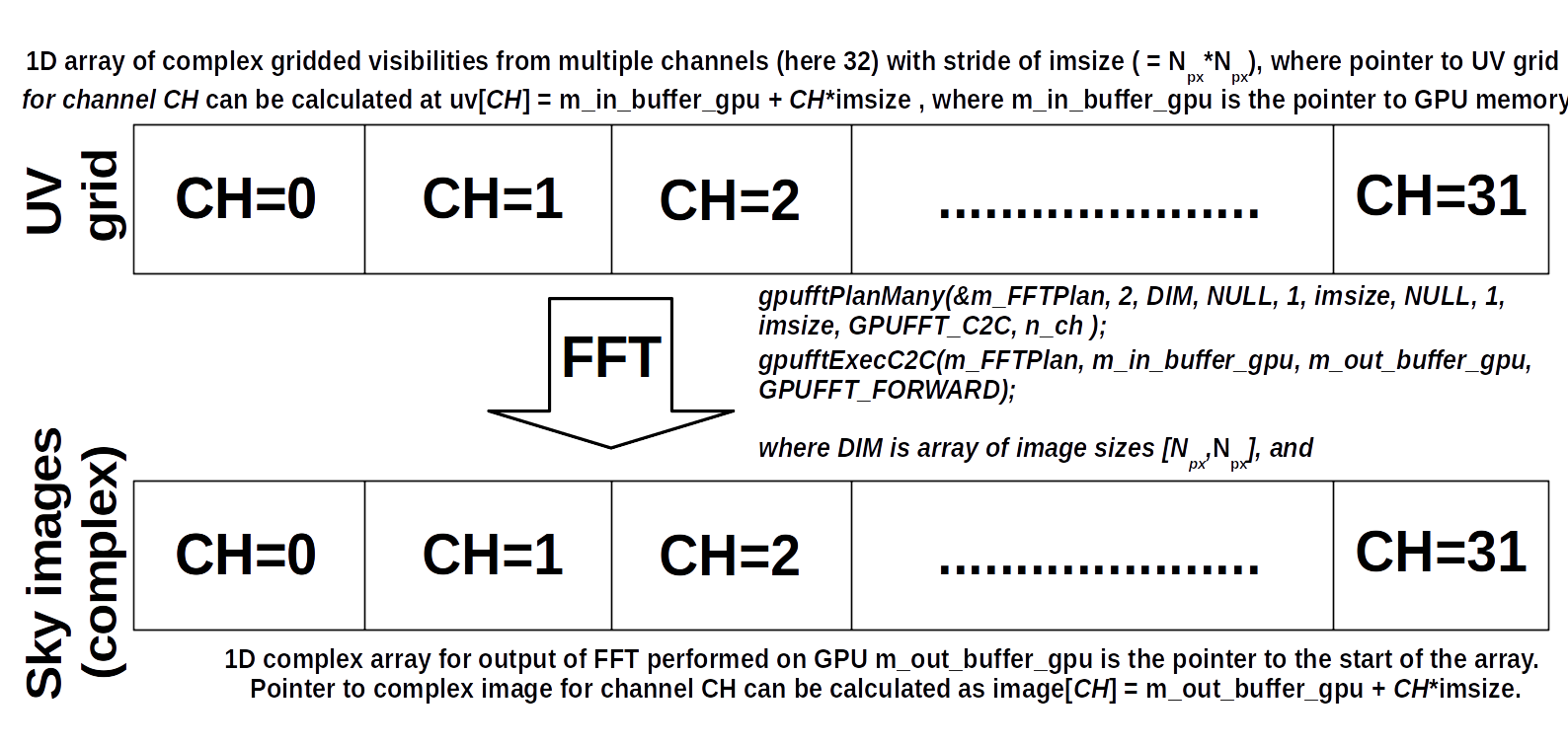} 
\caption{A diagram of the gridding and imaging of multiple channels ($N_{ch}$) performed on GPU. The input visibilities are gridded on a 2D UV grid ($N_{px}$, $N_{px}$) represented as a 1D array m\_in\_buffer\_gpu of complex floats. This is realised by a gridding kernel executed as shown in the listing in Figure~\ref{lst_layers_and_gridding}. This 1D array is used as the input to gpufftPlanMany and gpufftExecC2C functions to calculate FFT of all UV grids (for different frequency channels), and the final sky images can be obtained as real part of the complex images resulting from the FFT.}
\label{fig_gpu_gridding_imaging_diagram}
\end{center}
\end{figure*}

\subsubsection{Validation of the GPU imager}
\label{validation_gpu_images}

As mentioned earlier, the GPU imager was validated by comparing the resulting images with the images produced by its CPU version (its validation was described in Section~\ref{sec_cpu_imager_validation}). The EDA2 data leading to sky images in  Figure~\ref{fig_eda2_real_data_blink_casa_wsclean} were imaged with the CPU and GPU versions of the imager with 180, 1024 and 4096 pixels, and all the intermediate data products at various processing stages were compared. As a result, small differences ($\lesssim 0.1$\%) were found in the gridded visibilities. They translated into $\lesssim$0.5\% differences ($\lesssim $15\,mJy) in the final sky images in the areas of the sky where flux densities were >1\,Jy. This minimum flux density was required because calculation of relative differences in parts of the sky where flux density is $\approx$0 resulted in very high (over-estimated) ratios due to division by very small numbers. The corresponding differences for MWA data are typically much smaller than for EDA2 data. These discrepancies between GPU and CPU results are still being investigated. However, the current explanation is that they are caused by the imprecise nature of the floating point representation leading to non-commutative (although concurrency-safe) additions executed in no particular order by the gridding kernel. This is also the cause for the observed differences in the gridded visibilities between executions of the GPU imager. These effects are a small disadvantage of the current GPU version because results are not strictly reproducible. Thus, the test cases had to be modified in order to allow for a larger tolerance of the difference between the resulting sky images and the template images.

Nevertheless, in the future, we will try to eliminate these discrepancies by removing \texttt{atomicAdd} and implementing a deterministic algorithm based on the concept of parallel reduction \citep{HWU2023211}, which will lead to repeatable results in GPU version. This will also lead to code optimisation by removing \texttt{atomicAdd} calls, which are relatively costly operation due to being implemented as an active wait loop.
\begin{table*}[hbt!]
\begin{threeparttable}
\caption{A comparison of the execution times of gridding, imaging and both (total) for CPU and GPU implementations of the imager for a single time-stamp and frequency channel, tested for various image sizes using EDA2 and MWA data. The times were measured using function \texttt{std::chrono::high\_resolution\_clock::now()} on several different systems and GPUs used for during the development stage of the project (Table~\ref{tab_test_gpu_devices}).}
\label{cpu_gpu_1ft}
\begin{tabular}{ccccccc}
\toprule
\headrow 
\textbf{180 $\times$ 180}$^a$  & \textbf{1024 $\times$ 1024}$^a$ & \textbf{4096 $\times$ 4096}$^a$ & \textbf{456 $\times$ 456}$^a$ & \textbf{System} & \textbf{Processor} & \textbf{Operation} \\
 \textbf{(ms)} &  \textbf{(ms)} & \textbf{(ms)} & \textbf{(ms)} & &  & \\
\midrule
 5.8 & 106 & 2910 & 19.3 & Laptop A & CPU & Total \\
 2.7 &  16 &  209 &  3.4 &       & CPU & Gridding \\ 
 2.9 &  84 & 2610 & 14.7 &       & CPU & 2D FFTW \\ 
                          
\midrule
 15.7 & 14.0 & 13.0 & 16.3 & Laptop A  & GPU & Total \\
  0.1 &  0.1 &  0.1 & 0.05 &            & GPU & Gridding \\ 
 15.6 &  13.9 &  13.0 & 16.2 &         & GPU & 2D FFTW \\ 
\midrule
\midrule
 11.8 & 216.8 & 4769.7 & & Laptop B & CPU & Total \\
  5.8 &  31.9 &  440.4 & &       & CPU & Gridding \\ 
  5.5 & 174.7 & 4165.0 & &       & CPU & 2D FFTW \\ 
\midrule
 205.9 & 202.7 & 216.3 & & Laptop B & GPU & Total \\
   0.2 &   0.1 &   0.1 & &       & GPU & Gridding \\ 
 205.7 & 202.6 & 216.2 & &       & GPU & 2D FFTW \\ 
\midrule
\midrule
 7.7 & 144.6 & 2310.1 & & Workstation & CPU & Total \\
 3.8 &  23.2 & 310.7  & &       & CPU & Gridding \\ 
 3.5 & 113.0 & 1880.0 & &       & CPU & 2D FFTW \\ 
 \hline
 137.2 & 135.0 & 131.9 & & Workstation & GPU & Total \\
 0.1   &   0.1 &   0.1 & &       & GPU & Gridding \\ 
 137.1 & 134.9 & 131.8 & &       & GPU & 2D FFTW \\ 
\midrule
\midrule
 43.7   & 273.7 & 7550.1 & & Garrawarla & CPU & Total \\
  4.4   &  28.2 &  394.1 & &       & CPU & Gridding \\ 
 26.8   & 215.5 & 6921.1 & &       & CPU & 2D FFTW \\ 
 \midrule
 316.9  & 306.9  & 308.5  & & Garrawarla & GPU & Total \\
   0.05 &   0.05 &   0.04 & &       & GPU & Gridding \\ 
 316.8  & 306.8  & 308.4  & &       & GPU & 2D FFTW \\ 
\midrule
\midrule
 29.5 & 209.4 & 3613.0 & & Setonix & CPU & Total \\
  4.6 &  27.1 &  365.0 & &        & CPU & Gridding \\ 
 20.1 & 158.5 & 2944.8 & &        & CPU & 2D FFTW \\ 
  \midrule
 270.9 & 223.4 & 2427.5 & & Setonix & GPU & Total \\
   3.5 &   0.7 &    0.8 & &       & GPU & Gridding \\ 
 267.3 & 222.7 & 2426.7 & &       & GPU & 2D FFTW \\ 
 
\bottomrule
\end{tabular}
\begin{flushleft}
\begin{tablenotes}
\small
\item $^a$ Image sizes 180 $\times$ 180, 1024 $\times$ 1024 and 4096 $\times$ 4096 were tested with EDA2 data, while 456 $\times$ 456 with MWA data.
\end{tablenotes}
\end{flushleft}
\label{tab_cpu_vs_gpu}
\end{threeparttable}
\end{table*}

\subsubsection{Production version of the GPU Imager} 
\label{subsec_final_version}

The production version of the imager has to efficiently form images from multiple ($N$) frequency channels (tens to hundreds) and/or timesteps (tens to a hundred per second). This translates to multiple blocks of input visibility data that can be imaged in a parallel. We refer to these as \emph{layers} to distinguish them from the GPU programming concept of \emph{block} (of threads). These \emph{layers} correspond to multiple images, and their processing (gridding and FFT) is parallelised on GPU. A hierarchical approach to parallelisation will be considered in the future. The MPI standard may be used to distribute larger chunks of work across compute nodes; data are then processed in parallel on CPU cores and GPUs using OpenMP and HIP/CUDA, respectively. 

Visibilities resulting from previous steps (correlation and application of calibration) are already kept in GPU memory, and a continuous block of memory (of the size corresponding to N images, i.e. N$\times$N$_{px}$$\times$N$_{px}$ float values) is created to store the gridded visibilities and the resulting sky images. 

The gridding operation can be parallelised by processing different \emph{layers} in multiple GPU (CUDA or HIP) streams, which are execution queues of GPU kernel(s). Gridding kernels are scheduled for execution by adding them to these queues, and they are subsequently launched according to the available GPU resources (the maximum number of threads, memory, registers etc.). An alternative approach is to create multiple \emph{layers} (this is called grid in the CUDA nomenclature) of GPU blocks of threads with each layer corresponding to a separate UV grid of visibilities from a specific frequency channel or timestep. The same GPU gridding kernel is launched for multiple \emph{layers} by providing index \texttt{i} of the layer (i.e. UV grid) to be processed. The memory pointer to the specific UV grid (or sky image) in the input and output GPU memory can be calculated based on the index \texttt{i} of the sky image (frequency channel or timestep) as:

\begin{center}
\ttfamily
ptr[i] = ptr + i$\times$N$_{px}$$\times$N$_{px}\times$sizeof(float), 
\end{center}

where \texttt{ptr} is a pointer to allocated GPU memory and \texttt{i} is the image index. This is visualised in the listing in Figure~\ref{lst_layers_and_gridding}. Since, data from different frequency channels or times are processed independently there is no need for synchronisation mechanisms as kernels executed by different GPU streams or \emph{layers} access different parts of GPU memory.

The parallelisation with streams and \emph{layers} has been tested and benchmarked, and the \emph{layers} version was found to be about two times faster than the version using streams. The exact reason for this is unclear, but it may be caused by additional overhead of managing streams execution. At this point the gridding version using \emph{layers} of blocks is considered for the production version of the code.

\begin{figure*}[t]
\begin{sexylisting}{Multi-channel GPU gridding kernel and calling code}

// call to the GPU gridding kernel in file pacer_imager_multi_hip.cpp :
CPacerImagerMultiFreqHip::gridding_imaging_multi_freq
gridding_imaging_lfile_vis_blocks<<<nBlocks,NTHREADS>>>( ... m_in_cross_correlations_gpu, m_nCrossCorrBlockSize, ..., m_in_buffer_gpu, image_size, n_pixels, ...);

// GPU gridding kernel implementation in gridding_multi_image_cuda.cpp :
gridding_imaging_lfile_vis_blocks( ... gpufftComplex* in_visibilities_corr, int in_vis_corr_size, ...
gpufftComplex* out_visibilities_gridded, int image_size_cuda, ... )
{   
    // Calculating the required id 
    int i = blockDim.x * blockIdx.x + threadIdx.x;

    // different frequency channels are handled by different layers of blocks ( blockIdx.y ) :
    int freq_channel = blockIdx.y; // second block dimension means IMAGE BLOCK -> here frequency fine channel

    // calculate observing frequency based on freq_channel (known from Y coordinate of blockIdx)
    double freqMHz = first_channel_center_freq_mhz + freq_channel*channel_bw_mhz;
    double wavelength_cuda = SPEED_OF_LIGHT/(freqMHz*1e6); // c/freq_in_Hz

    // calculate pointer to output memory for frequency channel freq_channel:
    gpufftComplex* out_visibilities_gridded = out_visibilities_gridded_param + freq_channel*image_size_cuda;

    // obtain pointer to input visibility data (cross-correlations only here):
    gpufftComplex* baseline_data = get_cross_corr_data_antpol( ant1, pol1, cPol1, ant2, pol2, cPol2, inputs, n_inputs, n_channels, mapping_array, in_visibilities_corr  );

    float re = baseline_data[freq_channel].x;
    float im = baseline_data[freq_channel].y;

    // Beloe follows the code to place visibility value re/im on the UV grid (into the array  out_visibilities_gridded)
    ...
\end{sexylisting}
\caption{Listing of the GPU gridding code for multiple frequency channels. The GPU gridding kernel \texttt{gridding\_imaging\_lfile\_vis\_blocks} is called in the source file pacer\_imager\_multi\_hip.cpp, while the GPU gridding kernel itself is implemented in the source files gridding\_multi\_image\_cuda.h(cpp). }
\label{lst_layers_and_gridding}
\end{figure*}

\section{Benchmarking the GPU imager}
\label{sec_benchmarking}

The imager was tested and benchmarked on several compute architectures listed in Table~\ref{tab_test_gpu_devices}. As a first test, a single timestep and frequency channel were processed with the CPU and GPU versions of the BLINK imager, and the execution times obtained on different systems are summarised in Table~\ref{tab_cpu_vs_gpu}. In the next step, the current production versions of the code were benchmarked on Setonix using 600 100-ms timesteps and 32 channels for a total of 19200 EDA2 data points recorded on 2023-06-01 10:19:46 UTC. In this test 180x180, 1024x1024 and 4096x4096 images were produced using CPU and GPU versions of the code. Additionally, parallelisation of gridding using GPU streams and \emph{layers} was also benchmarked in the GPU version of the code. The results of this benchmarking are summarised in Table~\ref{tab_setonix_multichannel_benchmarking}.

\begin{table}
\caption{Results of benchmarking of multi-channel version of the code using CPU and two variations of GPU code. All tests were executed on Setonix on 1\,min of EDA2 visibilities in 100\,ms time resolution and 32 fine channels. The times in this table are mean and standard deviation of processing time of a single integration with 32 fine channels calculated based on 600 integrations.}
\vspace{-0.3cm}
\centering
\begin{scriptsize}
\begin{tabular}{@{}ccccc@{}}
\hline
\textbf{Version} & \textbf{Image} & \textbf{Gridding} & \textbf{2D FFT} & \textbf{Total} \\ 
                 & \textbf{size}  & \textbf{[ms]} & \textbf{[ms]} & \textbf{[ms]} \\ 
\hline
CPU & 180x180 & 150 $\pm$ 20 & 432 $\pm$ 40  & 768 $\pm$ 60 \\
    & 1024x1024 &  830 $\pm$ 20  & 4700 $\pm$ 200 & 6400 $\pm$ 300  \\
    & 4096x4096 &  11300 $\pm$ 700  & 8.7 $\pm$ 1.3 $\cdot10^4$ & 4 $\pm$ 0.5 $\cdot 10^3$ \\
\hline
GPU STREAMS & 180x180 & 4.4 $\pm$ 0.6 & 0.13 $\pm$ 0.04 & 4.5 $\pm$ 0.6 \\
    & 1024x1024 & 3.6 $\pm$ 0.5 & 1.74 $\pm$ 0.01 & 5.4 $\pm$ 0.5 \\
    & 4096x4096 & 4.5 $\pm$ 0.4  & 29.7 $\pm$ 0.03 & 34.2 $\pm$ 0.4  \\
\hline
GPU LAYERS & 180x180 & 0.024 $\pm$ 0.016 & 0.13 $\pm$ 0.03 & 0.17 $\pm$ 0.05 \\
    & 1024x1024 & 0.016 $\pm$ 0.007 & 1.98 $\pm$ 0.02 & 2.01 $\pm$ 0.014 \\
    & 4096x4096 & 0.024 $\pm$ 0.02 & 33.2 $\pm$ 0.04  & 33.25 $\pm$ 0.03  \\

\hline
\end{tabular}
\end{scriptsize}
\begin{flushleft}
\end{flushleft}
\label{tab_setonix_multichannel_benchmarking}
\end{table}

\begin{figure*}[t]
\begin{center}
\includegraphics[width=0.95\textwidth,angle=0]{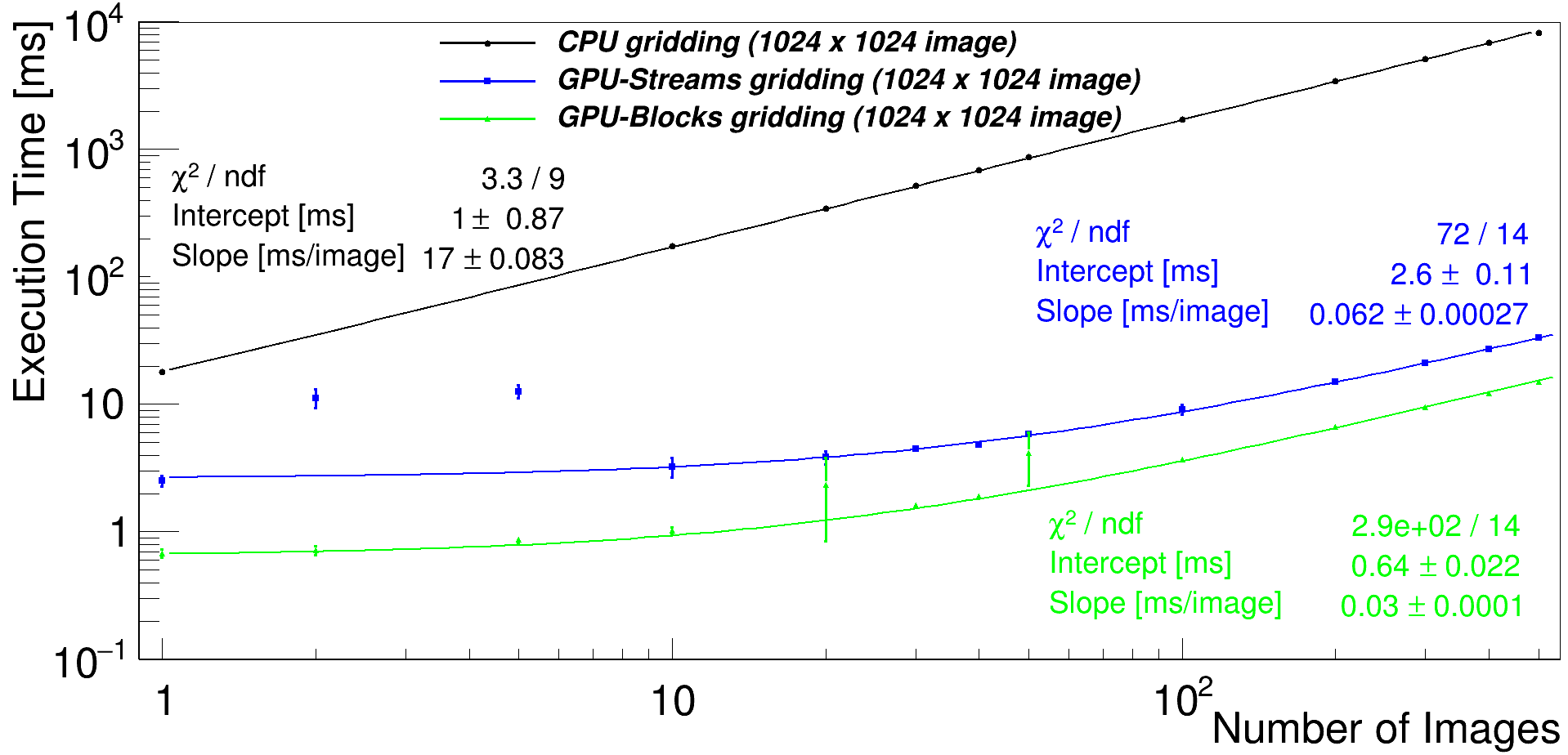} 
\caption{The execution time of gridding as a function of number of images using GPU streams (square points/blue fitted curve) and \emph{layers} of thread-blocks per image (triangle points/green fitted curve). The black data points and fitted curve are for CPU version of gridding, which confirms that it is between 1 and 2 orders of magnitude slower than its GPU implementation. The fitted parameters intercept (in ms) and slope (execution time per image) in ms are summarised in Table~\ref{tab_benchmark_fit_results}.}
\label{fig_gpu_gridding_benchmark}
\end{center}
\end{figure*}

\begin{figure*}[t]
\begin{center}
\includegraphics[width=0.95\textwidth,angle=0]{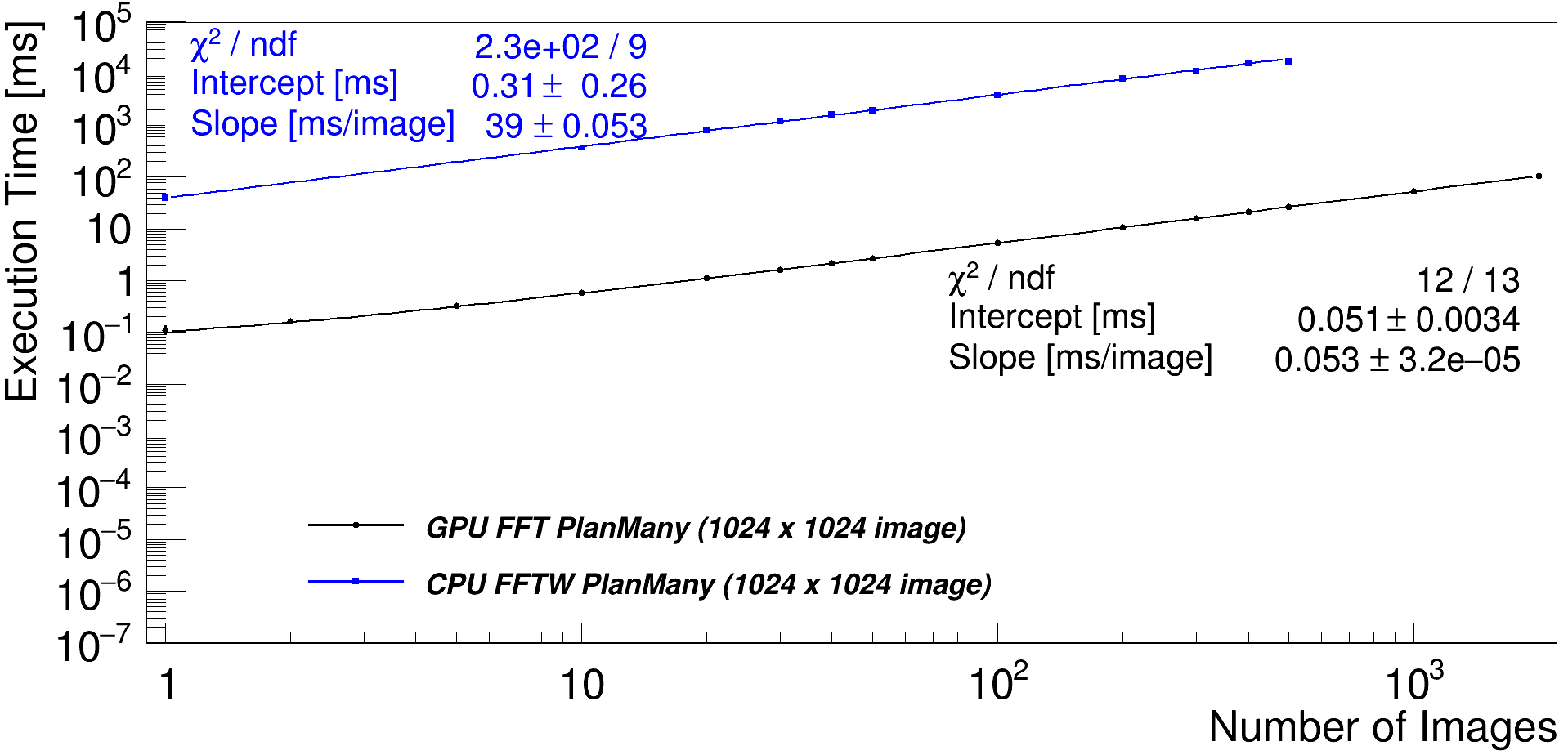} 
\caption{The execution time of 2D FFT using \texttt{gpufftw} (\texttt{hipfft} on Setonix) shown as black data points and fitted curve, and \texttt{fftw} library (blue data points and fitted curve). The black curve is about 1000 times lower, which means 2D FFT performed on GPU is about 3 orders of magnitude faster than its CPU counterpart. The fitted parameters, including the fitted slopes representing execution time per single image in ms, are summarised in Table~\ref{tab_benchmark_fit_results}. Based on the fitted slopes, it takes about 39.4\,ms per image in the CPU version vs. only 0.053\,ms (53\,usec) per image in the GPU version.}
\label{fig_gpu_fft_benchmark}
\end{center}
\end{figure*}

The production version of the multi-channel imaging code was also extensively benchmarked on Setonix.  The execution times of gridding and imaging as a function of number of images was measured for the same images sizes, and the results of these tests with GPU and CPU versions of the code are shown in Figures~\ref{fig_gpu_gridding_benchmark} and ~\ref{fig_gpu_fft_benchmark}. The execution times scale linearly with the number of images. Thus, the linear function was fitted to the data points and the fitted parameters (intercepts and slopes) are summarised in Table~\ref{tab_benchmark_fit_results}. The slope of the fitted line corresponds to processing time (gridding or 2D FFT) of a single image (i.e. units of ms/image). 

The slopes fitted in Figure~\ref{fig_gpu_gridding_benchmark} show that the GPU implementation takes only about 20, 30 and 60\,usec per image to perform parallel GPU gridding using \emph{layers} for 180x180, 1024x1024 and 4096x4096 images respectively. This is approximately 200, 580 and 3200 times faster than the CPU gridding for 180x180, 1024x1024 and 4096x4096 images respectively, which takes about 4.6, 17.1 and 195.1\,ms per image (slopes fitted to results of the CPU version). 

Similarly, the slopes fitted in Figure~\ref{fig_gpu_fft_benchmark} show that, in the GPU version, 2D FFT takes only about 1.7, 53.2, and 923.6 usec per image for 180x180, 1024x1024 and 4096x4096 images respectively. This is about 200, 740 and 1600 times faster than the CPU version for the respective image sizes. Overall, it is clear that GPU versions of gridding and 2D FFT are extremely fast (both take of the order of 83\,usec per 1024x1024-image) when they are performed in bulk on multiple images (in this case multiple frequency channels). They are between 2 and 4 orders of  magnitude faster than their CPU counterparts. 

Finally, the GPU version using GPU \emph{layers} (made of blocks of threads) is the most optimal GPU version about 2 times faster  with respect to GPU streams (Figure~\ref{fig_gpu_gridding_benchmark}). Additionally, the creation of GPU streams was measured to take between 15 and 20\,ms, but this is performed only once in the lifetime of the program and can be neglected. Therefore, clearly the most optimal version of the code is GPU version which uses GPU \emph{layers} to parallelise the gridding process, and this version is envisaged as the final production version of the imager for EDA2 and MWA data.


To determine the performance of both the gridding kernel and the 2D FFT routines, and whether better could be achieved, the code has been profiled with NVIDIA Nsight profiler (for the CUDA version) and the \texttt{rocprof} command (for the ROCm version). Code instrumentation (that is, placing timers around code regions) also gave informative insights.
What has been found for the gridding kernel is that execution times for the streams and \emph{layers} variants are very similar on the NVIDIA hardware, whereas streams perform poorly on AMD. The reason was found in the large initialisation time ($\sim12$\,ms) of the \texttt{hipStreamCreate} function call; the CUDA counterpart displays a much better performance ($< 0.1$\,ms). This is certainly due to ROCm framework being less mature than CUDA, also shown by some performance regressions encountered across versions of the framework. On the other hand, the \emph{layers} implementation in HIP outperformed the CUDA implementation on large problem instances, in terms of execution times.
The 2D FFT implementation is the one provided by the hardware vendor and included in the GPU programming framework (\texttt{cuFFT} for NVIDIA and \texttt{rocFFT} for AMD). With the FFT kernel's compute utilisation only slightly above $30\%$ and memory utilisation at $86\%$, the profiling of the CUDA version shows that the GPU FFT implementation is bound by the memory system. The reported memory bandwidth utilisation is $729.485$\,GB/s. Relative to the empirically measured peak memory bandwidth for the V100 GPU of $851.12$\,GB/s, the measured value for this kernel represents the $\sim85.7\%$ of the possible peak performance. Because of this, and because the MI100 GPU has a larger memory bandwidth than the V100, the ROCm implementation is $\sim30\%$ faster for the larger images (4096x4096).

%

   
\section{Summary and discussion}
\label{sec_summary}

We have presented a new GPU imager, which will be a part of a GPU-accelerated processing pipeline to search for FRBs and other transient phenomena with low-frequency telescopes. The primary objective is to search for these events in high-time resolution wide-field images from the MWA (e.g. SMART survey data) and all-sky images from SKA-Low stations as described in \citep{2022aapr.confE...1S}, which can yield even hundreds of FRB detections per year \citep{2024arXiv240104346S}. The main paradigm of the pipeline is to minimise I/O operations and process the data inside GPU memory. The full GPU pipeline is currently under active development, and will be described in an upcoming publication (Di Pietrantonio et al., in preparation).

The presented imager provides minimal functionality (``dirty images'') required by searches for bright radio transients, such as FRBs. It has been validated and tested on simulated visibilities from EDA2 and the MWA. It was also applied to real data from these telescopes and the resulting sky images were compared to images obtained with standard radio-astronomy packages (MIRIAD, CASA and WSCLEAN) using, as much as possible, the same imaging parameters (i.e. visibility weighting, angular size of pixels etc.).

A set of pre-compiler macros have been developed so that the code can be maintained in a single branch and compiled on any GPU architecture and programming framework (CUDA or HIP) regardless if the HIP framework is installed on the target system. During the development and testing, the imager was compiled, deployed and tested on several supercomputers (Setonix, Garrawarla and Topaz) at the Pawsey Supercomputing Centre as well as workstations and laptops with different NVIDIA GPU hardware (Table~\ref{tab_test_gpu_devices}). The speed-up of the GPU version can already be noticed even when  processing a single large (1024$\times$1024 or 4096$\times$4096) image (Table~\ref{tab_cpu_vs_gpu}). However, it becomes enormous for large amounts of data, for example 2D FFT performed on 500 images takes about 34.7\,ms per image using \texttt{fftw} library (CPU) and only 53.4\,usec per image (nearly 3 orders of magnitude faster) using cuFFT library (NVIDIA GPU) for 1024x1024 images. Similarly, for the same image sizes gridding takes about 16.45\,ms per image in the CPU version, and approximately 30\,usec per image (nearly 3 orders of magnitude faster) in the GPU version. Hence, on average, a single 1024x1024 image can be processed about 3 orders of magnitude faster on GPU than on CPU and gridding and imaging can be completed in about 80\,usec per image when processing is performed in bulk (on tens or hundreds of images).
Hence, real-time imaging of EDA2 or MWA is entirely feasible, and memory bandwidth remains the only limiting factor as discussed by Di Pietrantonio et al. (in preparation).

Nevertheless, there are still some areas for potential improvements. One of the most immediate optimisations is implementation of kernel executions on portions of data interleaved with memory transfers of other portions of data to/from the GPU memory. Furthermore, memory transfers can be further optimised by using pinned memory on the Host side\footnote{\url{https://developer.nvidia.com/blog/how-optimize-data-transfers-cuda-cc/}}. This will optimise the initial stage of the processing where the data are transferred from Host to GPU (device) memory. Such transfers are expected to be performed only in the beginning of the processing when the input data are read there from harddrives (off-line processing) or from the telescope's back-end (real-time processing). However, it is worth noting that some modern back-ends may be able to place the voltages from the telescope directly into GPU memory, and completely by-pass Host memory (either pinned or pageable). 

Finally, the all-sky GPU imager was already applied off-line to a few hours of EDA2 visibilities in 100\,ms time resolution to form all-sky images, and preliminary results of a pilot search for dispersed radio pulses (including FRBs) will be presented in a separate publication (Sokolowski et al., in preparation). The real-time imaging and FRB search system for EDA2 is currently under development. On the other hand, the incorporation of the imager into the full FRB search pipeline, and its application to a large sample of the MWA VCS data will be described in the upcoming publication (Di Pietrantonio et al., in preparation).

\begin{table*}
\caption{Parameters of straight lines fitted to execution times of gridding and 2D FFT as a function of number of images (Figures~\ref{fig_gpu_gridding_benchmark} and ~\ref{fig_gpu_fft_benchmark} respectively. The fitted slope corresponds to execution time per single image. Therefore, the slopes are larger for larger images. It can be seen that GPU gridding is about 2 orders of magnitude faster than CPU version for image sizes 180 and 1024, and about 3 orders of magnitude faster for the largest images (4096 pixels). Similarly, for the case of 2D FFT using GPU functions using \texttt{hipfft} library on Setonix when compared to FFTW CPU library. 2D FFT using \texttt{fftw} library led to very similar results when performed in 1 and 15 CPU threads. The optimal GPU version uses parallelisation of multiple frequency channels processing with GPU \emph{layers} (multiple blocks of GPU threads), and is about 2 times faster than parallelisation implemented with GPU Streams.}
\vspace{-0.3cm}
\centering
\begin{tabular}{@{}ccccc@{}}
\hline
\textbf{Function} & \textbf{Code}    & \textbf{Image} & \textbf{Fitted intercept} & \textbf{Fitted slope} \\
                  & \textbf{Version} & \textbf{Size}      & \textbf{[ms]} & \textbf{[ms/image]} \\
\hline
Gridding & GPU Streams & 180x180 & 2.327 & 0.05243 \\
         &             & 1024x1024 & 2.617 & 0.06162 \\
         &             & 4096x4096 & 2.148 & 0.09011 \\         
\hline
Gridding & GPU Layers & 180x180 & 0.6177 & 0.0234 \\
         &             & 1024x1024 & 0.6414 & 0.02962 \\
         &             & 4096x4096 & 0.2197 & 0.06162  \\         
\hline
Gridding & CPU         & 180x180 & -0.1093 & 4.628 \\
         &             & 1024x1024 & 1.004 & 17.08 \\
         &             & 4096x4096 & -0.1584 & 195.1 \\         
\hline
2D FFT   & GPU PlanMany & 180x180 & 0.05215  & 0.001722 \\
         &             & 1024x1024 & 0.05094 & 0.05321 \\
         &             & 4096x4096 & 0.08892 & 0.9236  \\         
\hline
2D FFT   & CPU FFTW\_PlanMany & 180x180 &  -0.01726 & 0.3396 \\
         &             & 1024x1024 & 0.3064 & 39.42 \\
         &             & 4096x4096 & -2.755 & 1481 \\         
\hline
\end{tabular}
\begin{flushleft}
\end{flushleft}
\label{tab_benchmark_fit_results}
\end{table*}

Although the imager has been originally developed and tested on the MWA and EDA2 data, the code is publicly available at \url{https://github.com/PaCER-BLINK-Project/imager} and it can be applied to other radio telescopes. 

\begin{acknowledgement}
This scientific work makes use of the Murchison Radio-astronomy observatory, operated by CSIRO. 
We acknowledge the Wajarri Yamatji people as the traditional owners of the Observatory site. 
Support for the operation of the MWA is provided by the Australian Government (NCRIS), under a contract to Curtin University administered by Astronomy Australia Limited. 
EDA2 is hosted by the MWA under an agreement via the MWA External Instruments Policy. This work was supported by resources provided by the Pawsey Supercomputing Research Centre’s Setonix Supercomputer (https://doi.org/10.48569/18sb-8s43), with funding from the Australian Government and the Government of Western Australia. Authors also acknowledge the Pawsey Centre for Extreme Scale Readiness (PaCER) for funding and support. 
 This work has also been co-funded and supported by ICRAR and Australian SKA Regional Centre (AusSRC). 
\end{acknowledgement}


\bibliography{pasa-sample}

\appendix

\end{document}